\documentclass[twocolumn,showpacs,amsmath,amssymb,prb,%
               superscriptaddress
               ,floatfix
              ]{revtex4}
\usepackage[T1]{fontenc}
\usepackage{times}
\usepackage{graphicx}
\usepackage{wasysym}
\usepackage{bm}

\begin{document}

\title{Impact of the inherent periodic structure 
       on the effective medium description \\
       of left-handed and related meta-materials}

\author{Th.\@ Koschny}
\affiliation{Ames Laboratory and Dept.~of Phys.~and Astronomy,
             Iowa State University, Ames, Iowa 50011, U.S.A.}
\affiliation{Institute of Electronic Structure and Laser, FORTH,
             71110 Heraklion, Crete, Greece}

\author{P.~Marko\v{s}}
\affiliation{Ames Laboratory and Dept.~of Phys.~and Astronomy,
             Iowa State University, Ames, Iowa 50011, U.S.A.}
\affiliation{Institute of Physics, Slovak Academy of Sciences, 
             845 11 Bratislava, Slovakia}

\author{E.~N.~Economou}
\affiliation{Institute of Electronic Structure and Laser, FORTH,
             71110 Heraklion, Crete, Greece}
\affiliation{Dept.~of Physics, University of Crete, 
             71110 Heraklion, Crete, Greece}

\author{D.~R.~Smith}
\affiliation{Dept.~of Electrical and Computer Engineering, Duke University,
             Durham, NC 27708, U.S.A.}
\affiliation{Dept.~of Physics, University of California, San Diego,
             La Jolla, CA 92093, U.S.A.}

\author{D.~C.~Vier}
\affiliation{Dept.~of Physics, University of California, San Diego,
             La Jolla, CA 92093, U.S.A.}

\author{C.~M.~Soukoulis}
\affiliation{Ames Laboratory and Dept.~of Phys.~and Astronomy,
             Iowa State University, Ames, Iowa 50011, U.S.A.}
\affiliation{Institute of Electronic Structure and Laser, FORTH,
             71110 Heraklion, Crete, Greece}

\date{November 2, 2004}

\begin{abstract}
We study the frequency dependence of the effective electromagnetic 
parameters of left-handed and related meta-materials of the split ring 
resonator and wire type. 
We show that the reduced translational symmetry (periodic structure)
inherent to these meta-materials influences their effective electromagnetic 
response. 
To anticipate this periodicity, we formulate a periodic effective medium 
model which enables us to distinguish the resonant behavior of 
electromagnetic parameters from effects of the periodicity of the structure. 
We use this model for the analysis of numerical data for the transmission 
and reflection of periodic arrays of split ring resonators, 
thin metallic wires, cut wires as well as the left-handed structures. 
The present method enables us to 
identify the origin of the previously observed resonance/anti-resonance
coupling as well as the occurrence of negative imaginary 
parts in the effective permittivities and permeabilities of those materials.
Our analysis shows that the periodicity of the structure can be neglected 
only for the wavelength of the electromagnetic wave larger than 30 space 
periods of the investigated structure.
\end{abstract}



\maketitle

\section{\label{section:intro} Introduction}

Recent progress in studies of left-handed meta-materials (LHM)\cite{Veselago68}
confirmed that the fabrication of structures with negative 
\textsl{effective} permittivity and permeability, and their application 
in technical praxis is possible.
The most promising structures are based on the combination of periodic arrays
of metallic split ring resonators (SRR) and thin metallic wires, 
a design proposed theoretically by Pendry \textsl{et al.} 
\cite{Pendry96b,Pendry98,Pendry99} 
and experimentally verified by Smith \textsl{et al.}
\cite{Smith00,Shelby01a,Shelby01b}.

It is assumed that in a well defined frequency interval both effective 
permittivity and permeability of LHM are simultaneously negative. 
Consequently, also the refractive index is negative \cite{SK}.
This theoretical prediction was supported experimentally by measurements 
of the transmission of the electromagnetic (EM) wave through the LHM:
A transmission peak was observed in the frequency region where 
the LH band is expected\cite{Smith00,Shelby01a}.
Negativeness of the index of refraction was verified experimentally
by the Snell's law experiment\cite{Shelby01b} and confirmed later by
other experiments
\cite{Parazzoli03,Houck03}.
Numerical simulations were performed which also observed a transmission 
peak in the resonant frequency interval\cite{Markos02,Markos02a,Markos02b}.
Effective electromagnetic parameters were calculated\cite{Smith02b}
by comparison of numerically obtained transmission and reflection amplitudes
of the LHM with theoretical formulas for a \textsl{homogeneous} slab. 
The obtained results confirmed that the refractive index of the LHM is indeed 
negative in the resonant frequency interval.  
Moreover, the obtained frequency dependence of the effective permittivity 
and permeability also agreed qualitatively with theoretical predictions. 
In particular, the effective magnetic permeability shows a resonant behavior,
\begin{equation}\label{one}
\mu(\omega)=
 1-\displaystyle{\frac{\omega_{mp}^2-\omega_{m}^2}{\omega_{}^2-\omega_{m}^2+
i\gamma\omega}},
\end{equation}
typical for lattice of SRR \cite{Pendry99} 
in the vicinity of the magnetic resonance frequency $\omega_m$.
The effective permittivity is determined by the electric response of the 
array of thin wires\cite{Sigalas95,Pendry98,Sarychev,Pokrovsky02b},
\begin{equation}\label{two}
\varepsilon(\omega)=
 1-\displaystyle{\frac{\omega_{p}^2}{\omega_{}^2+ i\gamma\omega}},
\end{equation}
and is negative if the frequency is smaller than the 
plasma frequency $\omega_p$.
Transmission data, obtained using either the transfer matrix method
\cite{Markos02a} or commercial software
\cite{Weiland01} was analyzed to find the dependence of the resonance 
frequency $\omega_m$ on the structural parameters of the SRR and on 
the design of the unit cell of the LHM structure. 

Further progress in numerical methods brought more accurate data
and strong evidence that the effective parameters of the LHM 
differ considerably from the theoretical prediction (\ref{one}, \ref{two}).
Although the main properties  -- resonant behavior of the
magnetic permeability at $\omega_m$ and negativeness
of the effective permittivity -- are clearly visible in 
the data, the effective medium picture is spoiled by
partially very significant anomalies: 

\smallskip
\noindent
\textsl{Resonance/anti-resonance coupling}. 
We expect the electric and magnetic response of the discussed 
meta-materials to be independent from each other. 
However, whenever there is a resonance in $\mathrm{Re}\ \mu$, 
we simultaneously observe an anti-resonant behavior in 
$\mathrm{Re}\ \varepsilon$\cite{OBrien02,Smith02b,Markos03a,OBrien04,Chen04}
and vice versa\cite{Koschny03b}. 

\smallskip
\noindent
The anti-resonant structures in the real part 
are accompanied by an negative
imaginary part\cite{Markos03a,Koschny03b}.

\smallskip
\noindent
\textsl{Misshapen, truncated resonances.} 
The divergence in $\mathrm{Re}\ n$ appears to be cut-off at the edges of the 
first Brillouin zone and, in particular, the negative regions of the magnetic
resonance in $\mu$ and cut-wire resonance in $\varepsilon$ 
do not return from large negative real part but seem to saturate in 
a rather shallow behavior.
The corresponding absorption peak in the imaginary parts 
is misshapen and highly asymmetric too\cite{Koschny03b}.

\smallskip
\noindent
\textsl{Discrepancy between $n$ and $z$ about the positions of the resonances.}
We expect the peaks (or zeros) in the index of refraction and the impedance 
to appear exactly at the resonance frequency. 
From the simulations, however, we find different frequencies 
from $n$ and $z$, respectively. 
This leads, for instance, to an ``internal structure'' of the magnetic resonance
as shown in Figs.~\ref{fig_srr_1} and \ref{fig_srr_2}. 
This structure can not easily be explained within the assumed effective 
medium picture\cite{Koschny03b}.

\smallskip
\noindent
\textsl{Additional spectral structures.}
Apart from structures around the anticipated contributions of the 
meta-material's constituents,
we observe a lot of additional structure, especially at higher 
frequency, which can not be accounted for. 

\smallskip
The above described observation, especially the negativeness of the imaginary
part of effective permittivity or permeability, raised objections
\cite{Depine04,Efros04b} of other groups. Efros\cite{Efros04b} argued that
the LHM can not be approximated by a homogeneous system because of the 
periodicity of the meta-material\cite{Pokrovsky02b,Efros04a}.

\smallskip
In this paper, we show that the observed artifacts in the homogeneous
effective approximation are quite generic. They are given by the
periodic structure of the investigated meta-materials. 
The periodic structure becomes important when the wavelength of the 
electromagnetic wave is comparable with the lattice structure of the 
material\cite{Koschny03b}. 
We proposed a more general description of the LHM, based on the concept 
of a \textsl{periodic effective medium} (PEM). 
This method enables us to distinguish between the resonant frequency 
dependence corresponding to Eq. \ref{one} and effects of the periodicity 
of the structure. 
We apply the PEM method for the analysis of numerical data  
obtained by the \textsl{transfer matrix method} (TMM).

\smallskip
The paper is organized as follows.

In Section \ref{section:hem} we first explain basic ideas of the 
\textsl{homogeneous effective medium} (HEM).
Special attention is given to the correction of the phase of 
the EM wave at the interfaces, which is crucial for any retrieval procedure.

In Section \ref{section:pem} we define and analyze 
one dimensional periodic structures.  
The analyzed medium consists of thin slabs of \textsl{homogeneous}
LH material separated by slabs of vacuum. 
We show that the approximation of such \textsl{periodic medium} by 
a \textsl{homogeneous} one give us effective parameters $\varepsilon$ and 
$\mu$ which possess unusual frequency dependences, similar to those observed
when we approximate meta-materials by a homogeneous medium.
This proves that the periodicity of meta-material must be taken into 
consideration in the analysis of the effective parameters.

The periodic effective medium is analyzed in two different formulations:
\textsl{continuous} (Section \ref{section:pem:cf}) and 
\textsl{lattice} (Section \ref{section:pem:lf}). 
The latter is more relevant for the analysis of numerical data since all 
known numerical algorithms use spacial discretization.

In Section \ref{section:results} we analyze transmission data, observed
from numerical simulations of periodic lattices of SRR, LHM and cut wires.  
We map these structures to periodic effective media which consist
of homogeneous slabs separated by vacuum. 
In this formulation, $\varepsilon$ and $\mu$ of the homogeneous slabs 
are free from any modifications of the resonant behavior. 
To show the role of the periodicity of the meta-materials more clearly, 
we also analyzed a lattice of SRR in which we filled the gaps of the SRR 
by a dielectric with very strong dielectric permittivity. 
This decreases the magnetic resonant frequency so that the wavelength 
of the incident EM wave is 25 times larger than the lattice period. 
We show that effective parameters again do not possess any deviations from 
resonant formula (\ref{one}).

A discussion of the applicability of various proposed models to the analysis 
of transmission data is given in Section \ref{section:discussion}. 
We discuss how the periodicity and anisotropy of the structure influence 
the transmission amplitudes and, subsequently, the effective parameters 
of the meta-materials.
Final conclusions are given in Section \ref{section:conclusion}.


\section{\label{section:hem} Homogeneous effective medium}

For the one-dimensional plain wave scattering problem at a homogeneous 
finite slab it is straightforward to obtain the scattering formulae. 
For the transfer matrices ${\bf T}_0$ for a single slice of vacuum and
${\bf T}_\mathrm{slab}$ for a single slice of homogeneous material 
with the thickness $d$ we find in wave-representation\cite{TMMDEF}
\begin{displaymath}
 \renewcommand{\arraystretch}{1.3}
 {\bf T}_0^{}(d)
 \ =\
 \left(\!
  \begin{array}{cc}
    e^{ikd} & 0  \\
    0 & e^{-ikd}
  \end{array}
 \!\right),
 \ 
 {\bf T}_\mathrm{slab}^{}(d)
 \ =\
 \left(\!
  \begin{array}{cc}
    \alpha(d) & \beta(-d) \\ 
    \beta(d)  & \alpha(-d)
  \end{array}
 \!\right)
\end{displaymath}
with the elements
\begin{eqnarray}
\alpha(d) &=& \cos(qd)\,+\,\frac{i}{2}\left(z+\frac{1}{z}\right)\sin(qd) \\
\beta(d)  &=& \frac{i}{2}\left(z-\frac{1}{z}\right)\sin(qd).
\end{eqnarray}
In the continuum formulation and for normal incidence 
the momentum $q$ inside the slab is related to the momentum $k$ in the vacuum 
by the index of refraction $n(k)=q/k$, the impedance $z$ is
defined by $z=\mu(\omega)k/q=q/(\varepsilon(\omega)k)$ for the 
TE and TM mode, respectively.
Here, $\mu(\omega)$ and $\varepsilon(\omega)$ denote the frequency-dependent
complex permeability and permittivity of the homogeneous medium.
On the lattice, ie.~when we are going to compare with TMM simulation results,
we have to take the modified dispersion relations $2-2\cos(k)-\omega^2=0$ 
in the vacuum 
and $2-2\cos(q)-\mu(\omega)\varepsilon(\omega)\,\omega^2=0$ inside the slab 
into account.
Then we have a modified $q=\mathrm{acos}\,(1-\mu\varepsilon(1-\cos\,k))$ 
which gets noticeable at higher frequencies.
Using the interrelation between the transfer matrix and the 
scattering matrix which defines the transmission ($t_\pm$) and 
refection ($r_\mp$) amplitudes,
\begin{equation}
 \label{scattering_S_and_T}
 \renewcommand{\arraystretch}{1.3}
 {\bf S}
 \ =\
 \left(\!
  \begin{array}{cc}
    t_+ & r_+ \\
    r_- & t_-
  \end{array}
 \!\right),
 \quad
 {\bf T}
 \ =\
 \left(\!
  \begin{array}{cc}
    t_{+}^{\vphantom{-1}}%
    \!-r_{+}^{\vphantom{-1}} t_{-}^{-1} r_{-}^{\vphantom{-1}}
    &
    r_{+}^{\vphantom{-1}} t_{-}^{-1}
    \\ 
    -\, t_{-}^{-1} r_{-}^{\vphantom{-1}} 
    &
    t_{-}^{-1}          
  \end{array}
 \!\right),
\end{equation}
we can calculate the transmission and reflection amplitudes for a sample composed of a left
vacuum slice of length $a$, followed by $N$ homogeneous unit cells of 
length $L$ in propagation direction, and terminated by a right vacuum slice 
of length $b$,
\begin{eqnarray}
\label{AD_tr_a}
t_{-}^{} &=&
\frac{\displaystyle e^{-i k N L}}%
     {\displaystyle \alpha(-d)\ e^{-ik(a+b)}} \\
\label{AD_tr_b}
r_{+}^{} &=&
 e^{-i k N L}\
 \beta(-d)\ e^{-ik(a-b)}\ t_{-}^{}\ .
\end{eqnarray}
In order to relate to the simulated scattering amplitudes computed numerically 
by the TMM by decomposition of the EM waves in the vacuum right of the sample 
with respect to the vacuum wave base left of the sample, it is convenient 
to introduce the normalized scattering amplitudes $T$ and $R$ which,
after $N$ unit cells, take the form 
\begin{eqnarray}
T &=& t_{-}^{}\, e^{ikNL} 
      \ =\
      \alpha^{-1}(-d)\ e^{ik(a+b)}
      \quad \\
R &=&
 \beta(-d)\ e^{-ik(a-b)}\ \ T.
\end{eqnarray}
In the continuum the scattering amplitudes of the homogeneous slab are
typically defined from interface to interface of the sample, ie.\@
assuming $a=b=0$. 
In the numerical simulation this is not possible because of the lattice:
we always have to make $1/2$ vacuum-transfermatrix step from the 
last left vacuum slice into the sample and another $1/2$ vacuum-transfermatrix
step out of the sample onto the first right vacuum slice.
Therefore, the TMM scattering amplitudes $T^{\,\textrm{\tiny(TMM)}}$
and $R^{\,\textrm{\tiny(TMM)}}$ are related to the normalized 
$T$ and $R$ involving an additional vacuum-phase compensation 
$T=e^{-ik}\,T^{\,\textrm{\tiny(TMM)}}$ and 
$R=e^{+ik}\,R^{\,\textrm{\tiny(TMM)}}$.

Now we can resolve the above scattering formulae with given amplitudes 
$T$ and $R$ obtained from the simulation (or measurement) of a meta-material
with respect to the material parameters impedance $z(\omega)$ and 
index of refraction $n(\omega)$. If the solutions are (virtually) independent
on the length of the sample those parameters define the
homogeneous effective medium (HEM) representation (or approximation) of the
respective meta-material. 
Then we have\cite{Smith02b}
\begin{eqnarray}
z_\mathrm{eff}(\omega) 
&=& \pm\,\sqrt{\frac{(1+R)^2-T^2}{(1-R)^2-T^2}}\,,
           \label{AD_inv_scattering_z} \\
n_\mathrm{eff}(\omega) 
&=& \pm\,\frac{1}{k L}
           \arccos\left(\frac{1-R^2+T^2}{2T}\right)\,+\,
           \frac{2\pi}{k L}\,m,\qquad
           \label{AD_inv_scattering_n}
\end{eqnarray}
with $m\in{\mathbb Z}$.
Note that we obtain $z_\mathrm{eff}$ and $n_\mathrm{eff}$ from the scattering 
amplitudes only up to a common sign and the real part of the effective index 
of refraction, $\mathrm{Re}\ n_\mathrm{eff}$ only as a residue class. 
The former issue can be resolved by imposing additional physical requirements,
for instance $\mathrm{Re}\ z\ge 0$ (causality).
The problem of the residue class for $\mathrm{Re}\ n_\mathrm{eff}$ can be 
addresses by considering different length $\{L_i\},\ i\in I\subset{\mathbb Z}$. 
Then we obtain a system of linear congruences, the solution of which
 -- if any -- 
is a reduced residue class modulo $2\pi/(k\,\mathrm{gcd}\ \{L_i\})$
given by the greatest common divisor of the lengths $L_i$.
Since due to the inherent periodic structure of real meta-materials in 
simulations and experiments the lengths of the sample can only be integral 
multiples of the unit cell's length, the minimum possible ambiguity for
$\mathrm{Re}\ n_\mathrm{eff}$ will be a residue class modulo $2\pi/(k L)$ 
where $L$ is the length of a single unit cell.
For physical reasons we can assume a smooth frequency dependence between 
resonances which enables us to obtain $\mathrm{Re}\ n_\mathrm{eff}(\omega)$
as the corresponding residue class of piecewise continuous functions. 
The correct branch then has to be chosen exploiting additional physical
information or assumptions of the model like the behavior of 
$n_\mathrm{eff}(\omega)$ at the plasma-frequency, in resonance induced 
transmission gaps an periodicity induced band gaps (discussed later). 
For known $n_\mathrm{eff}(\omega)$ and $z_\mathrm{eff}(\omega)$
the effective permeability $\mu$ and permittivity $\varepsilon$ can be 
defined as
\begin{eqnarray}
\mu_\mathrm{eff}(\omega) 
&=& 
 n_\mathrm{eff}(\omega)\,z_\mathrm{eff}(\omega), \\
\varepsilon_\mathrm{eff}(\omega)
&=&
 n_\mathrm{eff}(\omega)/z_\mathrm{eff}(\omega), 
\end{eqnarray}
respectively.

\medskip 
Results for the effective parameters of the HEM approximation of simulated
meta-materials like arrays of SRR or cut-wires, LHM and even multi-gap SRR
have been published by several authors\cite{%
Shelby01b,Markos02b,Smith02b,Koschny03b,OBrien02,OBrien04,Yen04}.
They all expose details which are in conflict with the simple effective medium
behavior in term of a resonant $\mu(\omega)$ and a plasmonic 
$\varepsilon(\omega)$, originally proposed by Pendry, even under the assumption
of an additional electric response of the SRR.
Typical examples are also shown in Figs.~\ref{fig_srr_1}, \ref{fig_srr_2},
\ref{fig_lhm_1}, and \ref{fig_lhm_2}.
All results show resonance/anti-resonance coupling in 
$\mu_\mathrm{eff}(\omega)$ and $\varepsilon_\mathrm{eff}(\omega)$ accompanied
by negative imaginary parts, apparently different resonance frequencies for
$n_\mathrm{eff}$ and $z_\mathrm{eff}$, the cut-off of the expected resonant 
positive (SRR) or negative (LHM) index of refraction, 
a misshapen, strongly asymmetric anticipated magnetic resonance in $\mu$ 
for the SRR and LHM or electric resonance in $\varepsilon$ for the cut-wire, 
and finally a lot of unexplained 
additional structure (erratic stop-band and passbands) at higher frequency.

Our extensive numerical simulations suggested that common cause for all these
problems has to be sought in the inherent periodicity, always present in
the artificial meta-materials as they are composed of a repetitions of a 
single unit cell. 
To prove that the behavior is generic and really independent on the details 
of the unit cell, and that we can reproduce each of the effect above purely as 
a consequence of periodicity in the propagation direction, 
we investigated the most simple model for an effective medium with a
non-trivial periodicity.

\section{\label{section:pem} Periodic effective medium}

\begin{figure}
\centerline{\includegraphics[width=8.5cm]{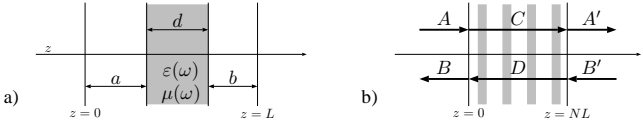}}
 \caption{%
  The layout of the single unit cell ($a$) and of a finite slab of the
  model periodic medium are shown. The shaded regions indicate the homogeneous
  core of the width $d$ which is characterized by the chosen appropriately
  model functions $\mu(\omega)$ and $\varepsilon(\omega)$, sandwiched by two
  vacuum slabs. $L$ is the length of a single unit cell, $N$ the number of 
  unit cells in the slab in propagations direction.
  Periodic boundary conditions apply in the directions perpendicular to 
  the propagation direction $z$.
 }
 \label{fig_model_1}
\end{figure}

To study the impact of the periodicity, or more precise the reduced 
translational symmetry of the sample in propagation direction, we consider 
a sample composed of a repetition of the unit cell shown in 
Fig.~\ref{fig_model_1}, finite in direction of propagation and infinite 
perpendicular to it. The unit cell consists of a thin homogeneous core
of thickness $d$ characterized by arbitrary $\mu(\omega)$ and 
$\varepsilon(\omega)$, sandwiched by two slabs of vacuum with thickness 
$a$ and $b$ which break translational invariance. $L$ is the length of one 
unit cell, $N$ the number of unit cells in propagation direction.
To make a connection to our meta-materials we choose a simple Lorentz-type
resonant form of $\mu(\omega)$ and/or $\varepsilon(\omega)$ to represent 
the magnetic and cut-wire response of the SRR. 
To model the LHM we would add a plasmonic term in $\varepsilon(\omega)$
to account for the response of the continuous wires.
Now we can calculate the scattering amplitudes for this model and subject 
them to the HEM inversion discussed in the previous section.
The description (or approximation) of the scattering amplitudes for a given 
meta-material in terms of the effective parameters of such a periodic medium 
as defined in Fig.~\ref{fig_model_1} 
will be denoted an "periodic effective medium" (PEM).

The following results will show that this periodic medium can expose all 
the problematic effect discussed above. 
In a subsequent section we shall then demonstrate that this also applies to 
the simulated real meta-material. Their effective behavior can be decomposed
into a "well-behaving" effective response of the resonances and a 
contribution of periodic structure described by the PEM.

\subsection{\label{section:pem:cf} Continuum formulation}

With the transfer matrices ${\bf T}_0$ and ${\bf T}_\mathrm{slab}$ 
introduced above, we can express the total transfer matrix of a finite slab 
of the periodic effective medium defined in Fig.~\ref{fig_model_1} 
in the form
\begin{displaymath}
 \left(\!
  \begin{array}{c}
   A' \\
   B'
  \end{array}
 \!\right)
 \ =\
 {\bf T}_0^{-1}(NL)\
 \Big[\ 
 {\bf T}_0^{}(b)\
 {\bf T}_\mathrm{slab}^{}(d)\
 {\bf T}_0^{}(a)\
 \Big]^N
 \
 \left(\!
  \begin{array}{c}
   A \\
   B
  \end{array}
 \!\right).
\end{displaymath}
As expected from the $z$-inversion symmetry both transfer matrices 
${\bf T}_0$ and ${\bf T}_\mathrm{slab}$ are unimodular, 
obviously is $\det{\bf T}_0=1$ and a short calculation
yields $\det{\bf T}_\mathrm{slab}=\alpha(d)\,\alpha(-d)+\beta^2(d)=1$.
Therefore we can easily calculate the $N$-th power of the unimodular 
2x2 matrix above by diagonalizing it and computing the $N$-th power of 
its eigenvalues\cite{Born-Wolf-Principles-of-Optics}.
Using the interrelation between the transfer matrix and the 
scattering matrix, we obtain the transmission and reflection amplitudes 
corresponding to those computed numerically by the TMM,
\begin{eqnarray}
\label{AD_tr_a2}
t_{-}^{} &=&
\frac{\displaystyle e^{-i k N L}}%
     {\displaystyle \alpha(-d)\ e^{-ik(a+b)}\ U_{N-1}(p)\ -\ U_{N-2}(p)}
\qquad\\
\label{AD_tr_b2}
r_{+}^{} &=&
 e^{-i k N L}\
 \beta(-d)\ e^{-ik(a-b)}\ U_{N-1}(p)\ \ t_{-}^{}\ .
\end{eqnarray}
Here, the $U_N=U_N(p)$ are the Chebyshev polynomials of the second kind,
$U_n(z)=\sin((n+1)\,\mathrm{acos}\,z)/(1-z^2)^{1/2}$, taken at the argument 
\begin{eqnarray}
\label{AD_tr_p}
\lefteqn{p \ =\ \cos(qd)\,\cos\big[k(L-d)\big]} && \\
&& \qquad -\
\frac{1}{2}\left(z+\frac{1}{z}\right)\sin(qd)\,\sin\big[k(L-d)\big]
\ .\ \nonumber
\end{eqnarray}
The wave vector $q=n(\omega)\,k$ and the impedance $z(\omega)$ 
refer to the homogeneous core of the unit cell.
For the normalized scattering amplitudes $T$ and $R$ after $N$ 
unit cells we find
\begin{eqnarray}
\label{AD_TR_a}
T &=& \left(
      \alpha(-d)\ e^{-ik(a+b)}\ U_{N-1}(p)\ -\ U_{N-2}(p)
      \right)^{-1} \quad \\
\label{AD_TR_b}
R &=&
 \beta(-d)\ e^{-ik(a-b)}\ U_{N-1}(p)\ \ T.
\end{eqnarray}


\begin{figure*}
\centerline{\includegraphics[width=17.cm]{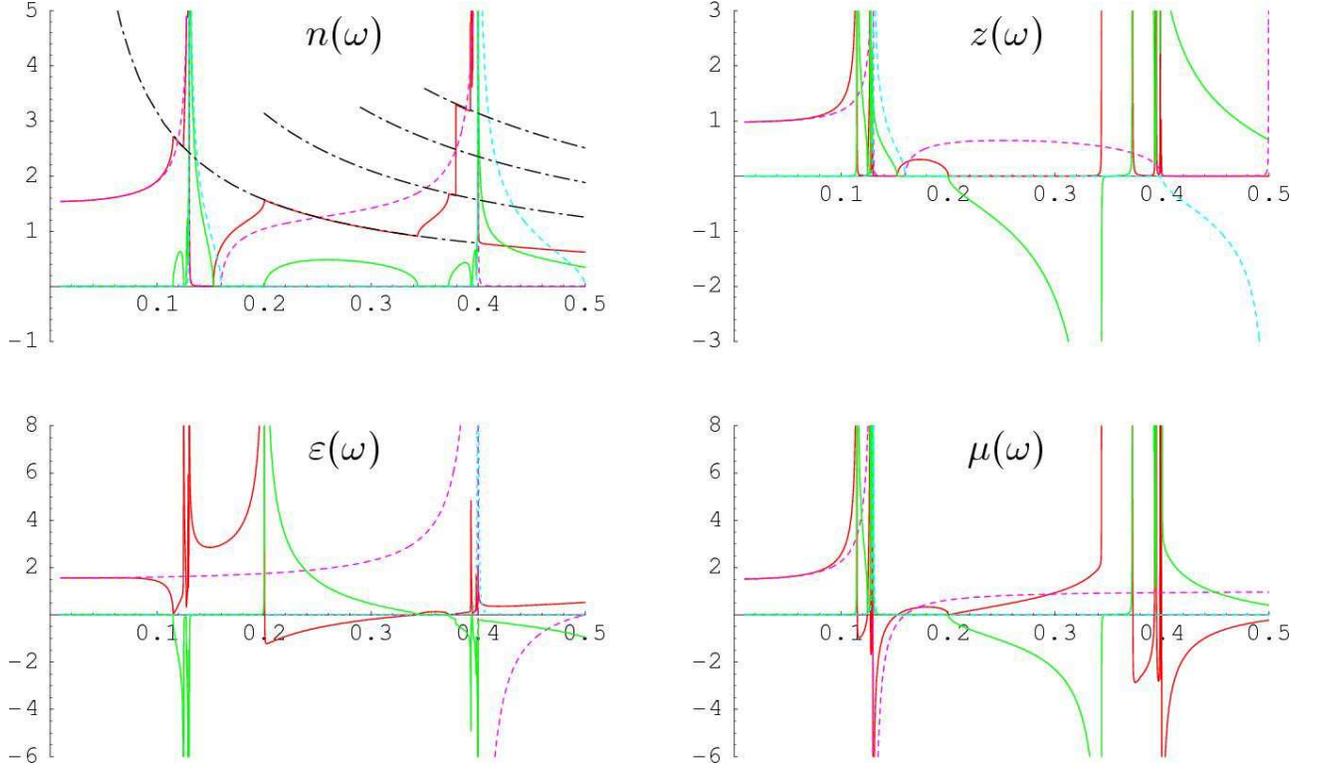}}
 \caption{%
  (color online) The HEM inversion 
  (Eqs.~\ref{AD_inv_scattering_neff_2}, \ref{AD_inv_scattering_zeff})
  of the analytic continuum PEM scattering amplitudes
  (Eqs.~\ref{AD_TR_a}, \ref{AD_TR_b}) for model SRR-type material parameters
  $\omega_m=0.13$, $\omega_{mp}=0.16$ for the magnetic,
  and $\omega_e=0.4$, $\omega_{ep}=0.5$ for the electric response
  and $\gamma=10^{-4}$ (cf.~Eqs. \ref{muH}, \ref{epsH}).
  The homogeneous core located in the middle of the unit cell was $d=L/10$ thick.
  The retrieved real (red, purple) and imaginary (green, turquoise)
  parts of effective parameters are shown as a function of frequency $\omega$.
  The dashed lines show the real (purple) and imaginary (turquoise)
  parts of the anticipated homogeneous parameters (Eqs.~\ref{muH}, \ref{epsH})
  and corresponding index of refraction and impedance.
  The dash-dotted black lines in $\mathrm{Re}\ n_\mathrm{eff}$ indicates the upper
  edge of the Brillouin zone, $n_\mathrm{edge}=k_\mathrm{edge}/k=m\,\pi/(k L)$.
 }
 \label{fig_ana_srr_cont}
\end{figure*}

Now we shall discuss what happens if we try to approximate the
explicitly periodic medium discussed above by a homogeneous effective medium.
This basically corresponds to our previous attempts to describe the 
periodic meta-materials by an homogeneous effective medium.
We have two options: First, we could simply consider the analytic scattering
amplitudes (\ref{AD_TR_a}, \ref{AD_TR_b}) derived above to be those of
a homogeneous system and try to solve for effective material parameters
$\varepsilon_\mathrm{eff}(N,\omega)$ and $\mu_\mathrm{eff}(N,\omega)$.
This has the advantage that the approximation can deal with a possible
residual length dependence of the approximate homogeneous medium, 
leaving an explicit possibility to assess the quality of the approximation.
The disadvantage is that we have to handle the rather complicated 
structure of the formulae arising from the Chebyshev polynomials.
The second approach is to assume that an exact correspondence of the
periodic effective medium to an homogeneous effective medium exists. 
This assumption is supported by the length-independence (after appropriate 
phase compensation) of the conventionally inverted simulation data. 
If there is such a homogeneous effective medium we can write the transfer
matrix of the periodic medium in terms of the transfer 
matrix for the homogeneous slab,
\begin{eqnarray}
 \label{AD_T_approx_a}
\lefteqn{%
 {\bf T}_0^{-1}(N L)\
 \Big[\
 {\bf T}_0^{}(b)\
 {\bf T}_\mathrm{slab}^{}(d)\
 {\bf T}_0^{}(a)\
 \Big]^N } \\
&=&
  {\bf T}_0^{-1}(N L)\
 ^\mathrm{(eff)}{\bf T}_\mathrm{slab}^{}(N L)\ ,
 \nonumber
\end{eqnarray}
which implies in particular for a system length of only a single unit cell
\begin{equation}
 \label{AD_T_approx_b}
 {\bf T}_0^{}(b)\
 {\bf T}_\mathrm{slab}^{}(d)\
 {\bf T}_0^{}(a)\
 \ =\
 ^\mathrm{(eff)}{\bf T}_\mathrm{slab}^{}(L)\ .
\end{equation}
Since for a homogeneous slab the identity
${\bf T}_\mathrm{slab}^{N}(L)={\bf T}_\mathrm{slab}^{}(N L)$
holds, finding a $^\mathrm{(eff)}{\bf T}_\mathrm{slab}^{}(L)$ that
satisfies equation (\ref{AD_T_approx_b}) in turn implies length independence
of the homogeneous effective medium description. 
Note that ${\bf T}_\mathrm{slab}(d)$ has only two independent elements,
because $\beta(d)=-\beta(-d)$ is anti-symmetric and the determinant 
$\alpha(d)\,\alpha(-d)+\beta^2(d)=1$ is fixed, such that we can 
calculate the matrix elements $\alpha(d)=(1-\beta^2(-d))/\alpha(-d)$
and $\beta(d)=-\beta(-d)$ from $\alpha(-d)$, $\beta(-d)$.
The assumption (\ref{AD_T_approx_b}) imposes a restriction on the 
boundaries of the periodic medium in propagation direction. 
The off-diagonal elements of ${\bf T}_\mathrm{slab}$ are anti-symmetric 
but on the left side of equation (\ref{AD_T_approx_b}) this symmetry is
broken by the phase factors $e^{ik(a-b)}$ and $e^{-ik(a-b)}$ introduced
in the off-diagonal elements by the two vacuum slabs.
As a consequence the description as a homogeneous medium is only possible
for $a-b=0$. Besides choosing a symmetric unit cell in the first place 
we may alternatively compensate the factor $e^{-ik(a-b)}$ in the 
reflection amplitude $R$ 
(which works simultaneously for all lengths, see equation \ref{AD_TR_b}), 
effectively redefining the boundaries of the system 
such that the slab is centered in the unit cells.
%
%
In terms of the (normalized) scattering amplitudes $T$ and $R$, 
for the single unit cell we then have the conditions
\begin{eqnarray}
  \alpha^{-1}(-d)\ e^{ik(L-d)}
 & = \quad T \quad = &
  \alpha_\mathrm{eff}^{-1}(-L) , 
  \label{AD_T_1}\\
  \beta(-d)\ T
 & = \quad R \quad = & 
  \beta_\mathrm{eff}(-L)\ T . \quad
  \label{AD_R_1}
\end{eqnarray}
We already know how to invert the right side of these equations, 
this is just what we did in the retrieval procedure for the HEM 
in the previous section. 
Defining renormalized scattering amplitudes $T'= T\,e^{-ik(L-d)}$ and
$R'=R\,e^{-ik(L-d)}$, we could apply the same procedure to the left side. 
Note that the possibility that we can always solve $T$ and $R$ for 
$\varepsilon_\mathrm{eff}(\omega)$ and $\mu_\mathrm{eff}(\omega)$
guarantees a solution of equation (\ref{AD_T_approx_b}).
In other words, there is always an exact, length independent description of 
the periodic effective medium as a homogeneous effective medium characterized
by $\varepsilon_\mathrm{eff}(\omega)$ and $\mu_\mathrm{eff}(\omega)$.
There is no freedom to chose the boundaries of the homogeneous medium 
relative to the periodic medium.
As shown above, we get the full information about the homogeneous
effective medium which describes a given periodic effective medium 
characterized by $n(k)$, $z(k)$ and the geometry $d$, $L$ already
from the first unit cell.
Inserting the renormalized transmission and reflection amplitudes 
(\ref{AD_T_1}, \ref{AD_R_1}) for a single unit cell into the inverted 
scattering formulae above we obtain
\begin{eqnarray}
 \cos\left(n_\mathrm{eff}(k)\, k L \right) &=&
 \cos(n k d)\cos\big[k (L-d)\big]
 \label{AD_inv_scattering_neff_1} \\
&&
 -\frac{1}{2}\left(z+\frac{1}{z}\right)
 \sin(n k d) \sin\big[k (L-d)\big]
 \nonumber
\end{eqnarray}
where $n(k)$ and $z(k)$ are the parameters of the material slab in the
middle of the unit cell of the periodic effective medium. 
With the $p$ defined in equation (\ref{AD_tr_p}) and $q=n(k)k$ 
we can write simpler
\begin{equation}
 n_\mathrm{eff}(k) \ =\
 \pm\frac{1}{k L} \arccos \Big[ p(n,z;k) \Big]  +\
           \frac{2\pi}{k L}\,m \
 \label{AD_inv_scattering_neff_2}
\end{equation}
with $m\in{\mathbb Z}$.
The problem with the signs of $n_\mathrm{eff}$ and $z_\mathrm{eff}$, 
as well as with the ambiguity of $\mathrm{Re}\ n_\mathrm{eff}$ is similar, 
and can be resolved the same way as for the case of the homogeneous slab 
discussed above.
Analogously we can express the impedance $z_\mathrm{eff}$ of the 
effective homogeneous medium in terms of the $n$ and $z$ of the 
homogeneous core as
\begin{equation}
z_\mathrm{eff}(k) \ =\
\pm
 \sqrt{
  \frac{2p^+ +\left(z-1/z\right)\sin(qd)}%
       {2p^+ -\left(z-1/z\right)\sin(qd)}}
\label{AD_inv_scattering_zeff}
\end{equation}
where $q=n(k)k$ and
\begin{eqnarray}
\lefteqn{ p^+\ =\
\cos(qd)\,\sin\big[k(L-d)\big] } && \\
&& \qquad \ +\
\frac{1}{2}\left(z+\frac{1}{z}\right)\sin(qd)\,\cos\big[k(L-d)\big]
\ .\ \nonumber
\end{eqnarray}

\medskip
The parameters of an effective homogeneous medium describing the periodic
material from Fig.~\ref{fig_model_1}, which have been obtained from the 
formulae (\ref{AD_inv_scattering_neff_2}) and (\ref{AD_inv_scattering_zeff}), 
are shown in Fig.~\ref{fig_ana_srr_cont} for a concrete example 
of SRR-type $\mu(\omega)$ and $\varepsilon(\omega)$.
For the homogeneous core in the middle of the unit cell we have chosen 
\begin{eqnarray}\label{epsmu}
\varepsilon(k) &=& 1+(L/d)\,[\varepsilon_H(k)-1], \\
\mu(k)         &=& 1+(L/d)\,[\mu_H(k)-1]
\end{eqnarray}
with model functions 
%
\begin{equation}\label{muH}
\mu_H(\omega) 
=
 1-\frac{\omega_{mp}^2-\omega_{m}^2}{\omega_{}^2-\omega_{m}^2+i\gamma\omega} 
\end{equation}
and
\begin{equation}\label{epsH}
\varepsilon_H(\omega) 
=
 1-\frac{\omega_{ep}^2-\omega_{e}^2}{\omega_{}^2-\omega_{e}^2+i\gamma\omega}
\end{equation}
%
to emulate the anticipated magnetic and electric\cite{Koschny04a} 
resonances of the SRR. For a LHM-type behavior we have to add the plasmonic 
response of the continuous wire in the permittivity,
\begin{eqnarray}\label{epsLHM}
\varepsilon_H(\omega) &=&
 1-\frac{\omega_{p}^2}{\omega_{}^2+i\gamma\omega}
  -\frac{\omega_{ep}^2-\omega_{e}^2}{\omega_{}^2-\omega_{e}^2+i\gamma\omega}.\quad
\end{eqnarray}
According to a simple effective medium picture, we would expect that we can
approximate a homogeneous unit cell characterized by $\mu_H(\omega)$ and
$\varepsilon_H(\omega)$ by concentration the magnetic and electric 
polarizations into the homogeneous core of the periodic medium.
Fig.~\ref{fig_ana_srr_cont} shows the actual effective impedance 
$z_\mathrm{eff}(\omega)$ and index of refraction $n_\mathrm{eff}(\omega)$
obtained via the HEM inversion of the periodic medium. 
%
%
Comparing with the expected effective medium behavior (dashed lines) we
clearly see the typical anomalies in the shape and positions of the
resonances, the same qualitative behavior as observed for real SRR 
meta-materials in the literature and our own previous work.
The effective parameters of the HEM approximation of our periodic medium
model show the resonance/anti-resonance coupling in 
$\mu_\mathrm{eff}(\omega)$ and $\varepsilon_\mathrm{eff}(\omega)$ 
together with the negative imaginary part 
$\mathrm{Im}\ \mu_\mathrm{eff}(\omega)<0$ around the magnetic resonance
frequency $\omega_m$, and also a very involved behavior close to the cut-wire
resonance $\omega_e$. 
The effective index of refraction is cut-off at the edge of the Brillouin 
zone which corresponds to the appearance of additional band gaps origination 
from the periodicity rather than from the underlying material properties.
The qualitative behavior presented in Fig.~\ref{fig_ana_srr_cont} is generic
for a wide range of parameters $\omega_m$, $\omega_{mp}$, $\omega_e$, 
$\ldots$ of the resonances and $L$, $d$ of the geometry. 
If the periodic medium model is used with only the electric resonance
or with an additional plasmonic term in $\varepsilon(\omega)$, it 
qualitatively reproduces the observed deviations from the expected plain
effective medium behavior published for the array of cut-wires and the 
LHM, respectively.

Although the curves show most of the discussed abnormalities in the 
HEM description of the SRR, the analytic description matches the simulation 
and inversion results for the real meta-material present in literature 
not in all aspects.
Clearly, there are problems very close to the resonance frequencies.
Instead of the divergence in the effective index of refraction $n_\mathrm{eff}$
being virtually cut off at the upper edge of the first Brillouin zone
as observed in the simulations of the actual SRR meta-material, 
the analytic description produces a series of consecutive band gaps at
the boundaries of the first and higher Brillouin zones and
a lot of structure in the imaginary part of $n_\mathrm{eff}$.
The same holds for the analytic description applied to the periodic effective
medium model of the LHM (not shown).
Here, we particularly miss the cutoff at the bottom of the negative 
$n_\mathrm{eff}$ region.
In either case the underlying lattice in the simulation starts 
to become visible. Since the lattice has a finite lattice constant it 
cannot support arbitrarily large momenta, such that we expect additional 
effects if the continuum momentum $q$ reaches the order of 
$\pi/a_\mathrm{lattice}$. 
In order to understand also the details of the retrieved HEM parameters 
in our simulation of real SRR and LHM meta-materials we have to take the 
discretization lattice of the employed TMM into consideration. 
To see the modification of the continuum results by the discretization lattice
we have to derive the scattering formulae for the periodic medium model 
on the lattice.

\subsection{\label{section:pem:lf} Lattice formulation}

We follow the TMM introduced for the Maxwell equations 
by Pendry\cite{Pendry92a,Pendry92b,Pendry94,Pendry96} in the formulation described by 
Marko\v{s} and Soukoulis\cite{Markos02a}.
The electric and magnetic field, together with the spatially dependent
material relative constants $\mu_\mathrm{rel}({\bf r})$ and 
$\varepsilon_\mathrm{rel}({\bf r})$ which define the meta-material,
are discretized on the bonds of mutually dual lattices $\{{\bf m}\}$ and
$\{\tilde{\bf m}\}$.
With the renormalized material constants 
$\varepsilon_i({\bf m})=
 i\omega\,\varepsilon_0\,\varepsilon_\mathrm{rel}({\bf m}+{\bf e}_i/2)$
and
$\mu_i({\bf m})=
 i\omega\,\mu_0\,\mu_\mathrm{rel}(\tilde{\bf m}-\tilde{\bf e}_i/2)$,
used throughout this section,
we can write the transfer matrix equations for a stratification in
$z$-direction for the two independent
components $i\in\{x,y\}$ of the electromagnetic field.
Using (quasi-)periodic boundary conditions in the $\bot z$ plane we can
introduce a Fourier representation of the fields with respect to this plane
defining an in-plane momentum ${\bf q}$. 
To derive a scattering formula corresponding to the continuum case considered 
in the previous section we restrict ourself to the most simple case of normal
incidence, ie.~zero in-plane momentum ${\bf q}=0$. 
Then the transfer matrix for normal incidence takes the form
\begin{eqnarray}
\lefteqn{%
\left(
 \begin{array}{c}
  {\bf E} \\
  {\bf H}
 \end{array}
\right)_{m_z+1} 
\ =\ } && \label{AD_lattice_T_1} \\
&&
\left(
 \begin{array}{cc}
 {\bf 1}  &  {\bf A}(m_z) \\
 {\bf B}(m_z+1)  &  {\bf 1}+{\bf B}(m_z+1) {\bf A}(m_z)
 \end{array}
\right)\,
\left(
 \begin{array}{c}
  {\bf E} \\
  {\bf H}
 \end{array}
\right)_{m_z}. \nonumber
\end{eqnarray}
The generally ${\bf q}$-dependent matrices $\bf A$ and $\bf B$ reduce to 
simple off-diagonal form, with the product ${\bf B}(m_z+1) {\bf A}(m_z)$ 
diagonal,
\begin{eqnarray}
{\bf A}(m_z) &=& 
\left(
 \begin{array}{cc}
  0         &  \mu_y(m_z) \\
 -\mu_x(m_z) &  0                          
 \end{array}
\right), \\
{\bf B}(m_z) &=&
\left(
 \begin{array}{cc}
 0           &  -\varepsilon_y(m_z) \\
 \varepsilon_x(m_z)  &  0                             
 \end{array}
\right),
\end{eqnarray}
such that the transfermatrix (\ref{AD_lattice_T_1}) factorizes, 
reordering the electromagnetic field vector in the form $(E_x,H_y,E_y,-H_x)^T$, 
into a two-fold degenerated block-diagonal structure
\begin{eqnarray}
\lefteqn{%
\left(
 \begin{array}{cc}
  E_x &  E_y  \\
  H_y & \hspace{-1ex}-H_x
 \end{array}
\right)_{m_z+1} =\ } && \\
&&
\left(
 \begin{array}{cc}
 1  &  \mu_y(m_z) \\
 \varepsilon_x(m_z\!+\!1)  &  1+\varepsilon_x(m_z\!+\!1)\mu_y(m_z)
 \end{array}
\right)\!
\left(
 \begin{array}{cc}
  E_x &  E_y  \\
  H_y & \hspace{-1ex}-H_x
 \end{array}
\right)_{m_z}\!\!. 
\nonumber
\end{eqnarray}
Without loss of generality we can restrict ourself to consider 
just the first polarization. 
We denote the single-polarization transfer matrix for the ${\bf q}=0$ modes
in the last equation $T(m_z)$. 
It is expedient to introduce the decomposition
\begin{eqnarray}
T(m_z) &=& \tau_\varepsilon(m_z+1)\,\tau_\mu(m_z) \\[1mm]
       &=&
 \left(
  \begin{array}{cc}
   1 &
   0 \\
   \varepsilon(m_z+1) & 1
  \end{array}
 \right)
 \left(
  \begin{array}{cc}
   1 &
   \mu(m_z) \\
   0 & 1
  \end{array}
 \right).
\end{eqnarray}
Further we can factorize the $\tau$ into a vacuum and a material contribution,
$\tau_\varepsilon
 \ =\
 \tau_{(\varepsilon-\varepsilon_\mathrm{vac})}
  \tau_{\varepsilon_\mathrm{vac}}
 \ =\
 \tau_{\varepsilon_\mathrm{vac}}
  \tau_{(\varepsilon-\varepsilon_\mathrm{vac})}$
related to the polarization
for the magnetic and analog for the electric field step.
Note the renormalized vacuum permittivity 
$\varepsilon_\mathrm{vac}=i\omega\varepsilon_0$ 
and permeability $\mu_\mathrm{vac}=i\omega\mu_0$.
As expected, $T(m_z)$ is unimodular.
Now we can easily find the eigensystem; 
the eigensystem of the vacuum transfer matrix defines the plain wave basis 
on the lattice which we use to define the scattering formalism.
Because of the unimodularity the two eigenvalues $\lambda=e^{\pm ik}$ 
are mutually reciprocal and for the propagating modes we are interested in 
on the unit circle, ie.~$k$ is real. 
We get the characteristic polynomial 
$\lambda^2-\lambda\,\big[2+\varepsilon_x(m_z+1)\,\mu_y(m_z)\big]+1$, 
hence 
$\,\cos\,k=1+\varepsilon_x(m_z+1)\,\mu_y(m_z)/2$.
The two signs of $k$ correspond to the right- and left-moving waves.
Note that $\varepsilon$ and $\mu$ implicitly contain the $\omega$ dependence.
To obtain the scattering matrix on the lattice 
we need the wave-representation of the total transfer matrix of a unit cell.
The right and left eigenvectors of $T(m_z)$ are distinct,
$R_\lambda(m_z) = 
 \big(1,\,(\lambda-1)/\mu_y(m_z)\big)^T$ and
$L_\lambda(m_z) = 
 \big(1,\,(\lambda-1)/\varepsilon_x(m_z+1)\big)^{*T}\!/(\lambda+1)^{*}$,
and satisfy the orthogonality relation 
$L^+_{\lambda_i}(m_z)\,R^{}_{\lambda_j}(m_z) = \delta_{ij}$.
Note that we applied the common normalization to the left-eigenvectors 
in order to normalize the electric field component of all right-eigenvectors 
to one.
This is required for a clean definition of the scattering amplitudes 
analog to the continuum case.
Further, the two right- and the two left-eigenvectors are linearly independent,
respectively.
Therefore we may group the two right- and the two left-eigenvectors of the
vacuum transfer matrix into the matrices
\begin{equation}
 {\bf L}_0^{\,+}
 \ =\
 \left(
  \begin{array}{cc}
   \displaystyle\frac{1}{e^{ik}+1} & 0 \\
   \rule{0pt}{7mm}
   0 & \displaystyle\frac{1}{e^{-ik}+1}
  \end{array}
 \right)
 \left(
  \begin{array}{cc}
   1 &
   \vphantom{\displaystyle\frac{1}{e^{ik}+1}}
   \displaystyle\frac{e^{ik}-1}{i\omega\varepsilon_0} \\
   1 &
   \rule{0pt}{7mm}
   \vphantom{\displaystyle\frac{1}{e^{ik}+1}}
   \displaystyle\frac{e^{-ik}-1}{i\omega\varepsilon_0} 
  \end{array}
 \right) 
\end{equation}
\begin{equation}
 {\bf R}_0^{}
 \ =\
 \left(
  \begin{array}{cc}
   \vphantom{\displaystyle\frac{1}{e^{ik}+1}}
   1 & 1 \\
   \vphantom{\displaystyle\frac{1}{e^{ik}+1}}
   \displaystyle\frac{e^{ik}-1}{i\omega\mu_0} &
   \displaystyle\frac{e^{-ik}-1}{i\omega\mu_0} 
  \end{array}
 \right),\\[2mm]
\end{equation}
where the eigenvalues $\lambda=e^{\pm ik}$ satisfy the vacuum dispersion 
relation 
$2-2\cos\,k +\mu_\mathrm{vac} \varepsilon_\mathrm{vac} = %
 2-2\cos\,k -\omega^2 \mu_0 \varepsilon_0 = 0$
for the vacuum wave vector $k$,
and use the projector ${\bf R}_0^{}\,{\bf L}_0^{+}$ to obtain the 
wave-representation of the total transfer matrix ${\bf T}_\mathrm{tot}$
of the finite system as
\begin{equation}
{\bf T}_\mathrm{tot}(k) \ =\
{\bf L}_0^{+}(k)\, {\bf T}_\mathrm{tot}\, {\bf R}_0^{}(k).
\label{AD_T_wave_1}
\end{equation}
Then we get the usual definition of the scattering amplitudes from
the correspondence between the scattering and the transfer matrix
given by equation (\ref{scattering_S_and_T}).


\begin{figure*}
\centerline{\includegraphics[width=17.cm]{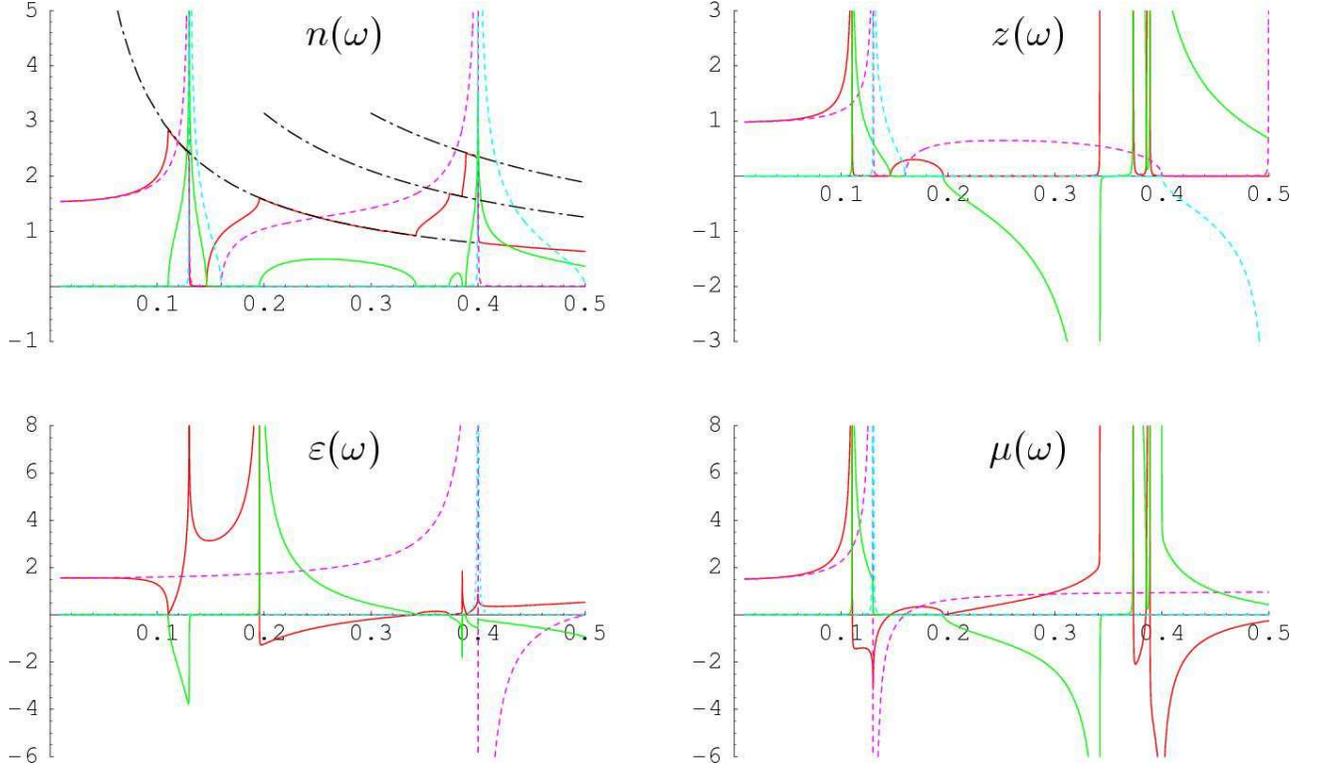}}
 \caption{%
  (color online) The HEM inversion 
  (Eqs.~\ref{AD_inv_scattering_z}, \ref{AD_inv_scattering_n})
  of the analytic lattice PEM scattering amplitudes (Eqs.~\ref{L_TR_new})
  for model SRR-type material parameters
  $\omega_m=0.13$, $\omega_{mp}=0.16$ for the magnetic,
  and $\omega_e=0.4$, $\omega_{ep}=0.5$ for the electric response
  and $\gamma=10^{-4}$ (cf.~Eqs. \ref{muH}, \ref{epsH}).
  The homogeneous core located in the middle of the unit cell was $d=L/10$ thick.
  The retrieved real (red, purple) and imaginary (green, turquoise)
  parts of effective parameters are shown as a function of frequency $\omega$.
  Note the reduction of the multiple band gaps seen in
  Fig.~\ref{fig_ana_srr_cont} around the the resonances to a single
  gap before each resonance.
  The dashed lines show the real (purple) and imaginary (turquoise)
  parts of the anticipated homogeneous parameters (Eqs.~\ref{muH}, \ref{epsH})
  and corresponding index of refraction and impedance.
  The dash-dotted black lines in $\mathrm{Re}\ n_\mathrm{eff}$ indicates the upper
  edge of the Brillouin zone, $n_\mathrm{edge}=k_\mathrm{edge}/k=m\,\pi/(k L)$.
 }
 \label{fig_ana_srr}
\end{figure*}


\begin{figure*}
\centerline{\includegraphics[width=17.cm]{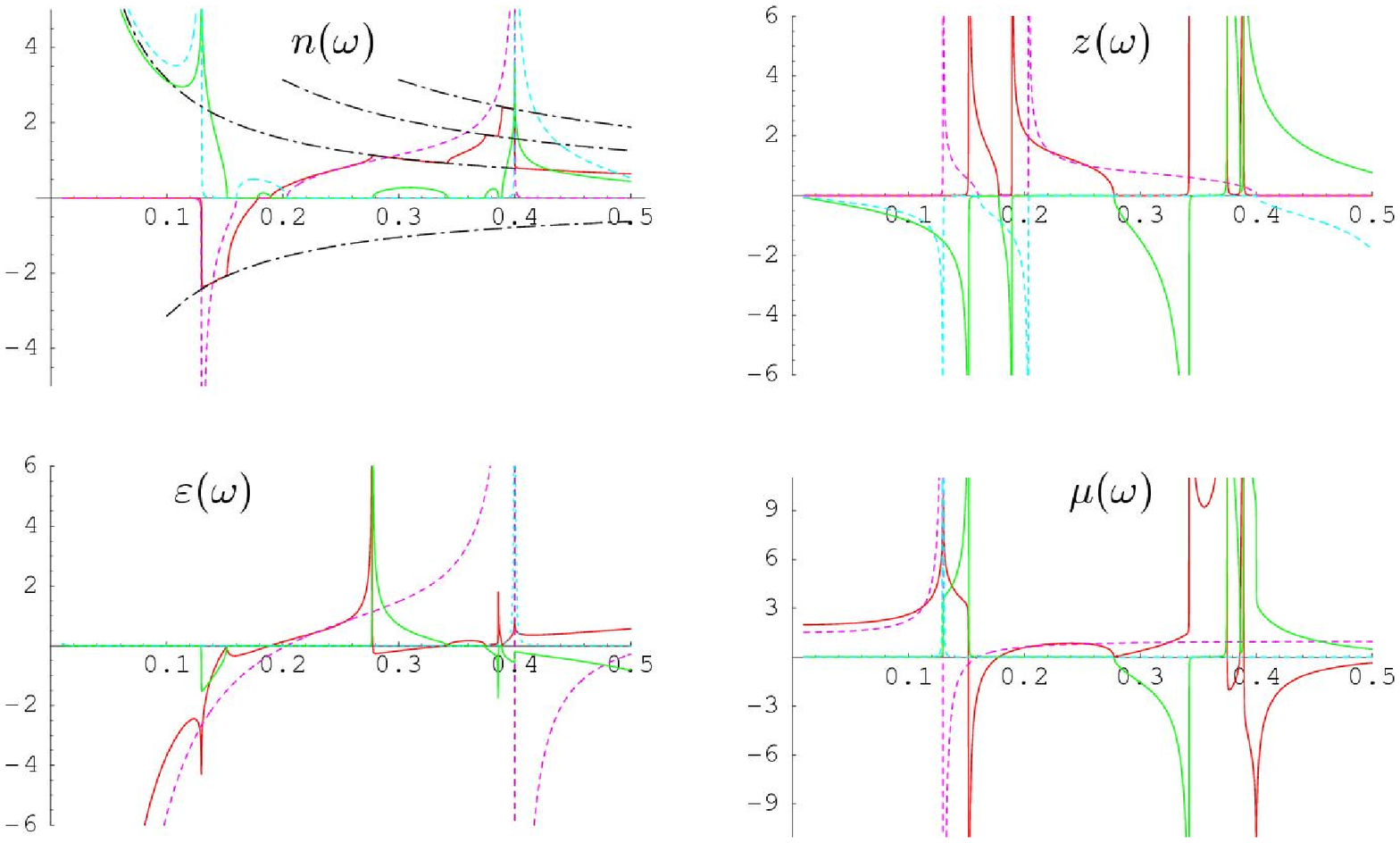}}
 \caption{%
  (color online) The HEM inversion 
  (Eqs.~\ref{AD_inv_scattering_z}, \ref{AD_inv_scattering_n})
  of the analytic lattice PEM scattering amplitudes (Eqs.~\ref{L_TR_new})
  for model LHM-type material parameters
  $\omega_m=0.13$, $\omega_{mp}=0.16$ for the magnetic,
  and $\omega_e=0.4$, $\omega_{ep}=0.5$, $\omega_p=0.27$ for the electric response,
  and $\gamma=10^{-4}$ (cf.~Eqs. \ref{muH}, \ref{epsLHM}).
  The homogeneous core located in the middle of the unit cell was $d=L/10$ thick.
  The retrieved real (red, purple) and imaginary (green, turquoise)
  parts of effective parameters are shown as a function of frequency $\omega$.
  The dashed lines show the real (purple) and imaginary (turquoise)
  parts of the anticipated homogeneous parameters (Eqs.~\ref{muH}, \ref{epsLHM})
  and corresponding index of refraction and 
  impedance.
  The dash-dotted black lines in $\mathrm{Re}\ n_\mathrm{eff}$ indicate the
  edges of the Brillouin zone, $n_\mathrm{edge}=k_\mathrm{edge}/k=m\,\pi/(k L)$.
 }
 \label{fig_ana_lhm}
\end{figure*}

\medskip
\paragraph*{The homogeneous slab.}
Now we have to consider the total transfer matrix of our meta-materials.
The most simple case is just a homogeneous slab of finite length.
%
On the lattice, the composition of the total
transfer matrix depends on the material discretization. 
We compute the total transfer matrix by starting from a right-eigenvector
of the vacuum base at the last vacuum site $m_z=0$ just before one side 
of the sample and apply successively the single-step transfer matrices
$T(m_z)$ until we reach the first site $m_z=n+1$ right of the sample 
for which the $T(m_z)$ is a vacuum step again. 
We have $n$ material layers inside the sample but $n+1$ transfer matrix step
which depend on the material parameters $\mu_y$, $\varepsilon_x$ of the sample.
Since we only have to consider a single polarization, we drop in the following
the $y,x$-indices in $\mu$ and $\varepsilon$ in order to improve readability.
Because in the discretized Maxwell equations the electric and magnetic 
fields live on mutual dual lattices, we distinguish three different 
single step $T(m_z)$ inside the sample instead of only one, as one would
expect for a homogeneous slab. $T(m_z)$ depends on $\mu(m_z)$ and 
$\varepsilon(m_z+1)$. Therefore the first step $T(0)$ inside the sample sees
only the electric response $\varepsilon(1)$ but no magnetic response of
the material. The subsequent steps see both, $\varepsilon$ and $\mu$, 
and are constant across the bulk of the sample. 
The last step back into the vacuum behind the slab is special again.
Both steps across the boundaries of the sample depend on the chosen 
material discretization. 
Here we adopt a symmetric material discretization\cite{Koschny04c}
which respects the $z$-isotropy such that the the steps into and out 
off the sample become equal.
Then we may calculate the wave representation as
\begin{eqnarray}
\lefteqn{%
T_\mathrm{tot}(k)
\ =\ 
{\bf L}_0^{+}\
 \tau_{\bar\varepsilon-\varepsilon} \,
 \big[\, \tau_\varepsilon \tau_\mu\,\big]^{n}\
 \tau_{\bar\varepsilon-\varepsilon_\mathrm{vac}} \,
 \tau_{\varepsilon_\mathrm{vac}}
 \tau_{\mu_\mathrm{vac}}\
 {\bf R}_0^{}
} &&
 \label{L_T_tot_new_2} \\
&=&
{\bf L}_0^{+}\,
 \tau_{\bar\varepsilon-\varepsilon_\mathrm{vac}}^{-1} \,
 {\bf R}
 \big[\, {\bf L}^{+} \tau_\varepsilon \tau_\mu\, {\bf R}\,\big]^{n}\
 {\bf L}^{+}\,
 \tau_{\bar\varepsilon-\varepsilon_\mathrm{vac}} \,
 T_\mathrm{vac}\,
 {\bf R}_0^{}
 \nonumber \\
&=&
 \big[
 {\bf L}^{+}
 \tau_{\bar\varepsilon-\varepsilon_\mathrm{vac}} 
 {\bf R}_0^{}
 \big]^{-1}
 \big[{\bf L}^{+} \tau_\varepsilon \tau_\mu {\bf R}\big]^{n}
 \big[
 {\bf L}^{+}
 \tau_{\bar\varepsilon-\varepsilon_\mathrm{vac}} 
 {\bf R}_0^{}
 \big]\,
 T_\mathrm{vac}(k)
 \nonumber
\end{eqnarray}
where $\{L_0^{+}, R_0^{}\}$ is the eigenbase of the vacuum transfer
matrix step $T_\mathrm{vacuum}$ with the eigenvalues $\lambda_0=e^{\pm i k}$
as before, but $\{L^{+}, R\}$ now denotes the eigenbase of the transfer matrix 
step $\tau_\varepsilon \tau_\mu$ inside the homogeneous medium with
the eigenvalues $\lambda=e^{\pm i q}$.
We made use of the afore mentioned identity 
$\tau_{\varepsilon_a}\tau_{\varepsilon_b}=\tau_{\varepsilon_a+\varepsilon_b}$.
The symmetric material discretization introduces the averaged
$\bar\varepsilon=(\varepsilon+\varepsilon_\mathrm{vac})/2$ 
at the material's surface.
As shown above, the wave vector in the vacuum $k$ and inside the homogeneous
slab $q$ satisfy the dispersion relations
$2-2\cos\,k+\mu_\mathrm{vac}\varepsilon_\mathrm{vac}=0$ and
$2-2\cos\,q+\mu\varepsilon=0$.
%
%
Since the matrix ${\bf L}^{+} \tau_\varepsilon \tau_\mu {\bf R}$ 
in Eq.~\ref{L_T_tot_new_2} is diagonal, 
we basically have to calculate the matrix
${\bf L}^{+} \tau_{\bar\varepsilon-\varepsilon_\mathrm{vac}} {\bf R}_0^{}$.
After some algebra we obtain for the homogeneous slab
\begin{displaymath}
 T_\mathrm{tot}(k) \ =\
 T_\mathrm{core}(k,n)\, T_\mathrm{vacuum}(k)
\end{displaymath}
with the diagonal $T_\mathrm{vacuum}(k)$ and
\begin{eqnarray}
\lefteqn{T_\mathrm{core}(k,n) \ =\ } && 
 \label{L_T_tot_new_3}
 \\[2mm]
 &&
 \frac{1}{\zeta}
 \left(
  \begin{array}{cc}
   \lambda^n\,G(\lambda) - \frac{1}{\lambda^{n}}\,G(\frac{1}{\lambda}) &
   (\lambda^n-\frac{1}{\lambda^{n}})\,(2-G(1)) \\[1mm]
   -(\lambda^n-\frac{1}{\lambda^{n}})\,(2-G(1)) &
   -\lambda^n\,G(\frac{1}{\lambda}) + \frac{1}{\lambda^{n}}\,G(\lambda)
  \end{array}
 \right)
 \nonumber
\end{eqnarray}
where 
\begin{eqnarray}
 G(\lambda) 
&=&
 \lambda\,\alpha(k)
 \ +\ \frac{\alpha(-k)}{\lambda}
 \ -\ \mu\varepsilon\,\alpha(k)\,\alpha(-k), 
 \\
\alpha(k)
&=&
 \frac{\varepsilon-\varepsilon_\mathrm{vac}}{2\varepsilon}
 \ +\
 \frac{\lambda_0(k)-1}{\varepsilon\mu_\mathrm{vac}},
 \\
 \zeta
&=&
 \frac{(\lambda-\lambda^{-1})(\lambda_0-\lambda_0^{-1})}
      {\varepsilon\mu_\mathrm{vac}},
\end{eqnarray}
with $\alpha(k)+\alpha(-k)=1$ and consequently 
$G(1)=1-\mu\varepsilon\,\alpha(k)\,\alpha(-k)$.
Further we have $\alpha(k)-\alpha(-k)=\zeta/(\lambda-\lambda^{-1})$
and $G(\lambda)=G(\lambda^{-1})+\zeta$.
Note the anti-symmetry of the off-diagonal elements.
Using again the definition of the scattering matrix (\ref{scattering_S_and_T}),
we find the transmission and reflection amplitudes as
\begin{eqnarray}
t_-(\omega,n) 
&=& \frac{\zeta\lambda_0}
         {\lambda^{-n}\,G(\lambda) -\lambda^n\,G(\lambda^{-1})}
\label{L_TR_new}
\\[1mm]
r_+(\omega,n) 
&=& 
 (\zeta\lambda_0)^{-1}\
 (\lambda^n-\lambda^{-n})\,(2-G(1))\ t_-(\omega,n).
 \nonumber
\end{eqnarray}
The non-vacuum factor of the lattice transfer matrix (\ref{L_T_tot_new_3}) 
appears to have the same symmetries as the transfermatrix of the 
homogeneous slab in the continuum:
the off-diagonal terms are anti-symmetric, the diagonal terms are mutual
complex conjugates if $\varepsilon_\mathrm{rel}$ and $\mu_\mathrm{rel}$ are 
real.

\medskip
\paragraph*{The periodic medium.}
Knowing the transfer matrix of the finite slab it is now easy to obtain 
the transfer matrix for a sample of multiple unit cells 
of the homogeneous as well as the periodic medium with the unit cell 
corresponding to Fig.~\ref{fig_model_1}$a$.
We can reduce the wave representation of the total transfer 
matrix to a product involving the wave representation of the homogeneous core
we already know and some additional vacuum transfer matrix steps
$T_0$ for the free space in the unit cell.
We assume the measures $a$, $b$ and $d$ in Fig.~\ref{fig_model_1}
to correspond to $n_a$, $n_b$ and $n_d$ layers on the lattice.
Then we get for the total transfer matrix of $N$ unit cells of the periodic
medium using 
$\tau_{\varepsilon_a+\varepsilon_b}=\tau_{\varepsilon_a}\tau_{\varepsilon_b}$
and consequently
$\tau_{\bar\varepsilon-\varepsilon}\,
 \tau_{\bar\varepsilon-\varepsilon_\mathrm{vac}}=1$
\begin{eqnarray}
\lefteqn{T_{\mathrm{pm},N}(k) } && 
 \label{L_T_pm_new_1} \\ [1mm]
&=&
 {\bf L}_0^{+}
 \Big[\,
 {(
  \tau_{\varepsilon_\mathrm{vac}}
  \tau_{\mu_\mathrm{vac}}
 )}^{n_b}
 \big[\,
 \tau_{\bar\varepsilon}
 \tau_{\mu}\,
 (\tau_\varepsilon
 \tau_\mu)^{n_d-1} \, 
 \tau_{\bar\varepsilon}
 \tau_{\mu_\mathrm{vac}}
 \,\big] 
 \hspace{10mm} \nonumber \\
& & \hspace{10mm}
 {(
 \tau_{\varepsilon_\mathrm{vac}}
 \tau_{\mu_\mathrm{vac}}
 )}^{n_a-1}
 \Big]^N
 \tau_{\varepsilon_\mathrm{vac}}
 \tau_{\mu_\mathrm{vac}}
 {\bf R}_0^{}
 \nonumber \\
&=&
 \Big[\,
 {T_\mathrm{vac}}^{n_b}(k)\
 T_{\mathrm{core}}(k,n_d)\ 
 {T_\mathrm{vac}}^{n_a}(k)\, 
 \Big]^N\,
 {T_\mathrm{vac}}^{}(k).
 \nonumber 
\end{eqnarray}
with the $T_{\mathrm{core}}(k,n_d)$ defined in equation (\ref{L_T_tot_new_3}).
Since the phase factors $\lambda_0^{-(n_b-n_a)}$ and $\lambda_0^{n_b-n_a}$
introduced by the two vacuum slabs in the bracket on the last line
of equation (\ref{L_T_pm_new_1}) do explicitly break the anti-symmetry 
of the off-diagonal elements that is present for the single homogeneous 
slab in the continuum and, in the symmetric material discretization, 
also on the lattice, we can obtain a representation of the periodic medium
by a homogeneous medium only for the case $n_b=n_a$. 
As already explained for the continuum case this is
not a real restriction but instead just fixes the definition of the 
effective boundaries of the periodic medium.
In the numeric simulation we have to explicitly compensate the 
corresponding vacuum phases in the scattering amplitudes. 
We can use the Chebyshev formula to explicitly calculate the $N$-th power 
such that we get the transmission and reflection amplitudes for the periodic 
medium after $N$ unit cells in propagation direction as
\begin{eqnarray}
\lefteqn{t_-(\omega,N) \ =\ } && \label{L_TR_pm_new_a} \\
&& \lambda_0
   \left(
         {\frac{
           \lambda^{-n_d}\,G(\lambda) -\lambda^{n_d}\,G(\lambda^{-1})
          }{\zeta\,\lambda_0^{n_b+n_a}}\ 
          U_{N-1}-U_{N-2} }
   \right)^{-1},
 \nonumber
 \\[2mm]
\lefteqn{r_+(\omega,N) \ =\ } && \label{L_TR_pm_new_b} \\[1mm]
&& 
 \frac{
   \big(\lambda^{n_d}-\lambda^{-n_d}\big)\,\big(2-G(1)\big)
   \ U_{N-1}\ t_-(\omega,N)}
 {\zeta\,\lambda_0^{n_a-n_b}\ \lambda_0^{} }.
 \nonumber
\end{eqnarray}
where the argument of the Chebyshev polynomials $U_N(p)$ is given by
\begin{eqnarray}
\lefteqn{p \ = \
\cos(q n_d)\cos\big[k(n_a+n_b)\big]} && \\
&&
\qquad \ -\ \frac{G(\lambda) + G(\lambda^{-1})}{\zeta}\,
\sin(q n_d)\sin\big[k(n_a+n_b)\big].
\nonumber
\end{eqnarray}
As for the continuum formulation, we actually get all the information about 
the meta-material from the single unit cell.
Comparing the scattering amplitudes (\ref{L_TR_pm_new_a}, \ref{L_TR_pm_new_b})
on the lattice with the normalized scattering amplitudes for homogeneous slab 
in the continuum tells us how to do the phase compensation for the lattice-tmm
results: $T=\lambda_0^{-1}\,t_{-}$ and $R=\lambda_0^{-(n_b-n_a)}\,r_{+}$.
The condition for $T$ arises from the additional vacuum step $T_0$ into the
slab on the lattice, the compensation in $R$ results from the symmetric
definition of the boundary of the unit cell which is required to describe
the periodic by a homogeneous medium as explained above.

\medskip
\paragraph*{Continuum HEM inversion.}
Again we ask whether the model periodic medium from Fig.~\ref{fig_model_1}
can be represented by an effective homogeneous medium. 
Here we have two choices: 
~($i$) we can compare the scattering amplitudes of the lattice periodic 
medium with the scattering formulae (\ref{AD_T_1}, \ref{AD_R_1}) derived 
for the homogeneous slab in the continuum, or we can 
~($ii$) compare with the lattice scattering formulae for the homogeneous 
slab derived in this section. 
Moreover, we have to decide which material discretization to use.
In this paper we will concentrate on comparing the lattice scattering 
results to the continuum scattering formulae for the homogeneous slab, 
as we previously did with the standard inversion procedure to obtain 
effective $\varepsilon_\mathrm{eff}(\omega)$ and $\mu_\mathrm{eff}(\omega)$ 
from the meta-material simulations. 

Analytically, the effective material parameters obtained from the 
HEM inversion for the lattice formulation of the model periodic media used 
in the last section to emulate the SRR and LHM meta-materials are shown 
in Fig.~\ref{fig_ana_srr} and Fig.~\ref{fig_ana_lhm}, respectively.
As expected, the qualitative behavior is very similar to that found with
the continuum formulation. 
All the problematic effects seen in the previously published simulations,
like resonance/anti-resonance coupling, negative imaginary parts, deformed 
resonances, bad gaps and so on, are present.
The major difference to the continuum formulation becomes visible around the
resonances. Where we previously found a series of tiny periodicity band gap
around the resonances, in the lattice formulation we obtain a much simpler
structure with basically one gap before each resonance.
This is in excellent agreement with the numerical simulations, 
hence, expectedly, the lattice formulation compares much better to numerical
simulations also obtained via discretization of the Maxwell equations
than the continuum formulation.
The discussion of further details we shall defer to a dedicated section below.

\begin{table}
\begin{tabular*}{8.5cm}{l@{\hspace{3mm}}p{6.5cm}}
\hline

HEM      & Homogeneous Effective Medium, a homogeneous medium 
           characterized by $\mu(\omega)$ and
           $\varepsilon(\omega)$ which, 
           substituted for a finite meta-material slab,
           length-independently reproduces (or approximates)
           the given scattering amplitudes.
           Here always used in continuum formulation.
           Finding a HEM for given $T$, $R$ is called HEM inversion (if exact) 
           or HEM approximation. \\
 
PEM      & Periodic Effective Medium, a most simple periodic model-medium
           defined by $\mu_\mathrm{core}(\omega)$,
           $\varepsilon_\mathrm{core}(\omega)$ and a geometry
           shown in Fig.~\ref{fig_model_1} which length-independently reproduces
           (or approximates) given scattering amplitudes.
           Also used a priori with given $\mu_\mathrm{core}(\omega)$,
           $\varepsilon_\mathrm{core}(\omega)$ to demonstrate effects of the 
           periodicity. Here used in lattice formulation. \\

HEM(PEM) & The HEM which reproduces the scattering amplitudes
           calculated analytically from a given PEM. \\

\hline
\end{tabular*}
\caption{%
 \label{tab_acronymes}
 Summary of the effective medium related acronyms.
}
\end{table}

\section{\label{section:results} Simulation results}

In this section we now present actual TMM simulation results for real SRR
and off-plane LHM meta-materials.
All numerical simulation are done using an implementation of the TMM method
described by Marko\v{s} and Soukoulis\cite{Markos02a}. 
The meta-materials are uniformly discretized on a cubic lattice using a 
symmetric material discretization. 
The dimensions of the unit cell are 6x10x10 mesh steps, the single-ring SRR 
is a square ring of 7x7 mesh steps with a gap in the top side 
one mesh step wide. Propagation is for all cases along the SRR plane with 
the polarization of the incident plain wave such that the electric field
is parallel to the two continuous sides of the SRR. 
Therefore we have only magnetic coupling to the magnetic resonance of 
the SRR\cite{Smith00,Katsarakis04a}.
Periodic boundary conditions apply to both directions perpendicular to 
the direction of propagation.
For the off-plane LHM we add a one mesh-step thick continuous wire in front 
of the SRR such that the position of the wire is symmetric in the middle 
between two periodic repetitions of the SRR plane and centered with respect
to the gap in the SRR. The direction of the wire is parallel to the continuous
sides of the SRR, thus parallel to the incident electric field. 
All components of the meta-materials, the ring ring of the SRR and the 
continuous wire, are made from metal characterized by a constant relative 
permittivity of $\varepsilon_\mathrm{metal}=(-3.0+5.88\,i)\,10^5$ and
$\mu_\mathrm{metal}=1$. 
Note that the results do not depend much on $\varepsilon_\mathrm{metal}$ 
as long as it does not fall below a certain threshold\cite{Markos02a}.
The chosen value is reasonable to emulate metals like Cu, Ag, Au in the range 
of GHz to a few THz.
The rest of the unit cell is vacuum, there are no dielectric boards.
The special geometry of the unit cell has been carefully chosen to preserve 
the inversion symmetry of the unit cell in the two directions perpendicular 
to the direction of propagation.  
This allows us to consider the scattering for only one polarization 
as it avoids complications by cross-polarization terms in the 
scattering amplitudes\cite{Koschny04c}.
In this paper, we concentrate our consideration on the region around 
the magnetic resonance frequency $\omega_m$, where we expect 
$\mu_\mathrm{eff}(\omega)$ to become transitionally negative, 
for two reasons:
first this is the region of interest for any left-handed application, 
and second, this is the frequency window for which simulation data is
typically shown in the literature. 
A more detailed investigation of the higher frequency region, particularly 
the vicinity of the of the electric cut-wire response of the SRR and the 
intermediate periodicity band gaps will be published elsewhere.

In the following we show HEM inversion results for the scattering data 
numerically obtained for the meta-material with the TMM.
After the correct vacuum-phase compensation described above the inverted
HEM scattering formulae (\ref{AD_inv_scattering_z}, \ref{AD_inv_scattering_n}) 
are applied to the simulated $T$ and $R$ for meta-material slabs with a
thickness of one, two and three unit cell in propagation direction.
We shall denote the results as
$n_\mathrm{hem}(\omega)$ and $z_\mathrm{hem}(\omega)$
or
$\mu_\mathrm{hem}=n_\mathrm{hem}\, z_\mathrm{hem}$ and
$\varepsilon_\mathrm{hem}=n_\mathrm{hem}/z_\mathrm{hem}$,
correspondingly.
This approach is the same as chosen in the literature.
Then we find the PEM approximation for the simulated meta-material using the
lattice formulation for the analytic scattering formula of a model periodic
medium consisting of a homogeneous core which is a single discretization 
mesh-step thick and located the unit cell in the plain of the SRR gaps and 
the LHM continuous wire.  This constitutes the lattice equivalent of a 
single scattering plain in the continuum.
A model periodic medium, characterized by effective material constants
$\mu(\omega)$ and  $\varepsilon(\omega)$ 
of the homogeneous core, which reproduces the simulated $T$ and $R$ 
independent on the system length is called a periodic effective medium.
The numeric inversion of the lattice scattering formulae 
(\ref{L_TR_pm_new_a}, \ref{L_TR_pm_new_b}) is applied to the simulated 
$T$ and $R$ for the firsts unit cell of each meta-material, 
providing us with effective material constants 
$\mu_\mathrm{core}(\omega)$ and  $\varepsilon_\mathrm{core}(\omega)$ 
for the homogeneous core of the PEM approximation.
From the core parameters we can derive two further sets of effective 
parameters.
First, we calculate the HEM inversion of the PEM scattering data obtained 
from the retrieved 
$\mu_\mathrm{core}(\omega)$ and $\varepsilon_\mathrm{core}(\omega)$
and compare the results with the HEM inversion of the direct simulation data
to assess the quality of the PEM approximation. We denote this as HEM(PEM).
Second, we introduce the material parameters 
$\mu_\mathrm{pem}(\omega)$ and  $\varepsilon_\mathrm{pem}(\omega)$
of a homogeneous unit cell 
that would correspond to the PEM approximation in the effective medium limit,
equating the total electric and magnetic polarizations of the respective
unit cells,
\begin{equation}
\label{R_mu_pem}
\mu_\mathrm{pem}(\omega) 
\ =\ 1+\frac{n_d}{n_a+n_d+n_b}\ \big(\mu_\mathrm{core}(\omega)-1\big)
\end{equation}
and
\begin{equation}
\label{R_eps_pem}
\varepsilon_\mathrm{pem}(\omega) 
\ =\ 1+\frac{n_d}{n_a+n_d+n_b}\ \big(\varepsilon_\mathrm{core}(\omega)-1\big).
\end{equation}
The idea of this definition is to obtain parameters which we can compare
with those of the HEM inversion, becoming equivalent with the latter 
if we can truly neglect the periodicity of the material.
This allows us, to some degree, to consider the meta-material's 
electromagnetic response as being composed of an actual contribution of the
internal geometry of the meta-materials constituents 
and an explicit contribution of the periodic arrangement.


\begin{figure*}[p!]
\centerline{\includegraphics[width=14cm]{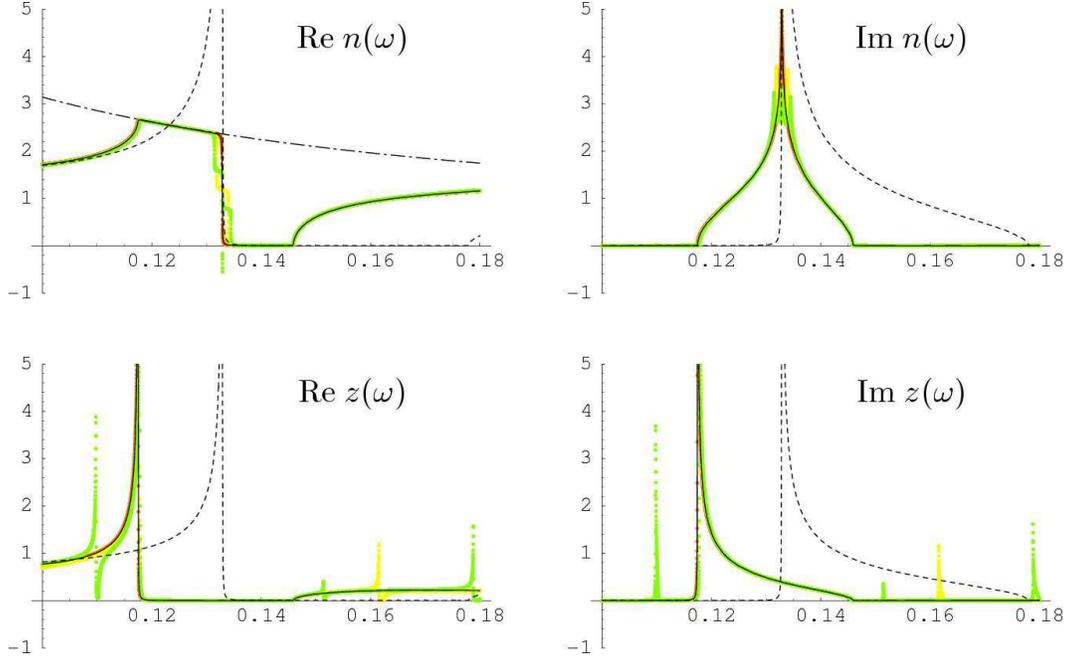}}
 \caption{%
 (color online) For the simulated SRR meta-material the effective index of refraction 
 $n_\mathrm{eff}(\omega)$ and impedance $z_\mathrm{eff}(\omega)$ are shown.  
 The colored curves represent the HEM approximation
 (Eqs.~\ref{AD_inv_scattering_z}, \ref{AD_inv_scattering_n})
 of the simulation data for the first 3 unit cells,
 the solid black line the HEM(PEM) approximation,
 and the dashed line the PEM approximation (cf.~Eqs.~\ref{R_mu_pem}, \ref{R_eps_pem})
 obtained from the first unit cell data.
 Note the different positions of the resonance for 
 $n_\mathrm{eff}(\omega)$ and $z_\mathrm{eff}(\omega)$.
 }
 \label{fig_srr_1}
\end{figure*}

\begin{figure*}
\centerline{\includegraphics[width=14cm]{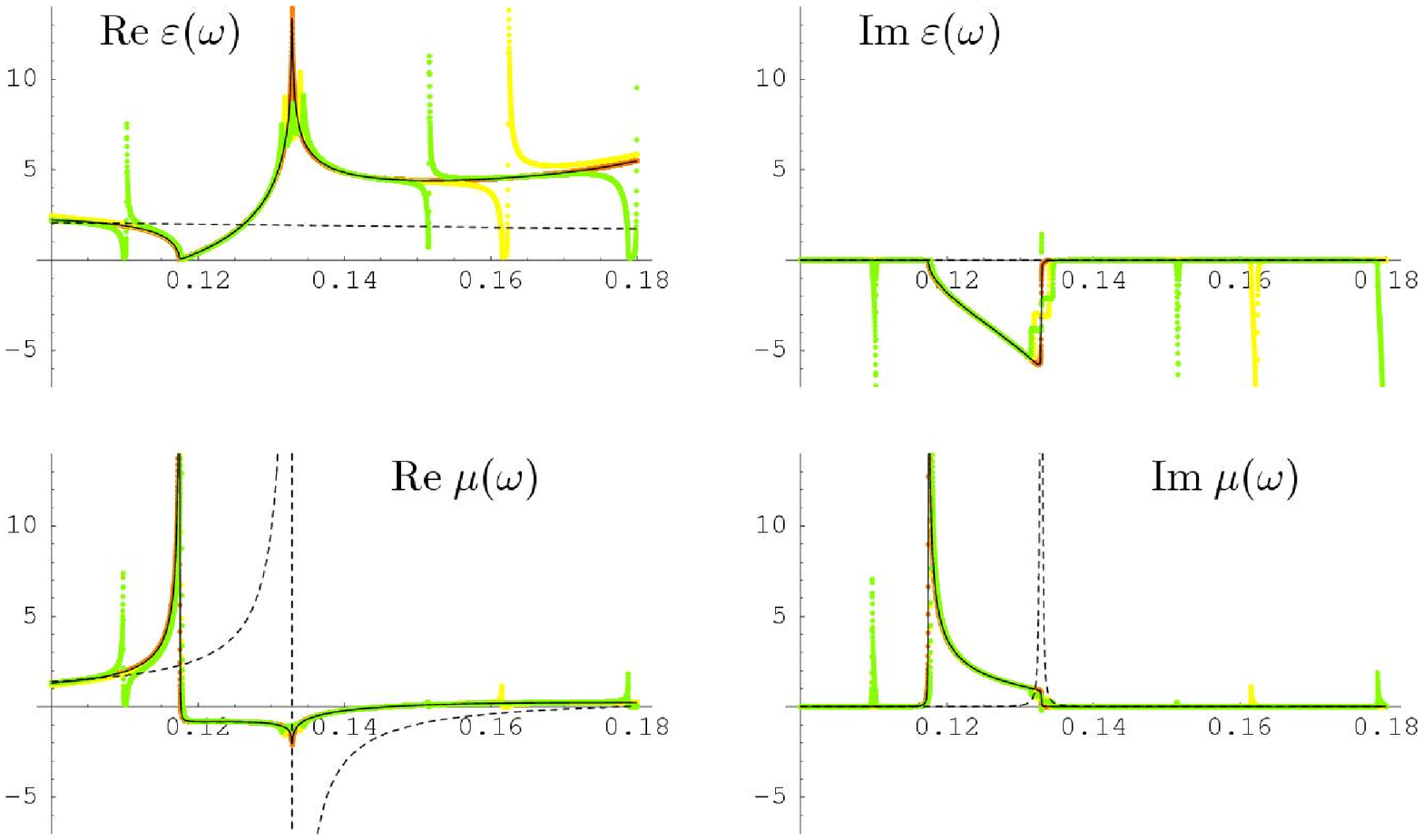}}
 \caption{%
 (color online) For the simulated SRR meta-material the effective permittivity
 $\varepsilon_\mathrm{eff}(\omega)$ and permeability $\mu_\mathrm{eff}(\omega)$
 are shown.  
 The colored curves represent the HEM approximation
 (cf.~Eqs.~\ref{AD_inv_scattering_z}, \ref{AD_inv_scattering_n})
 of the simulation data for the first 3 unit cells,
 the solid black line the HEM(PEM) approximation,
 and the dashed line the PEM approximation (Eqs.~\ref{R_mu_pem}, \ref{R_eps_pem})
 obtained from the first unit cell data.
 Note the anti-resonant behavior of the permittivity and the misshapen magnetic
 resonance in the frequency interval where $n_\mathrm{eff}(\omega)$ reaches the
 edge of the Brillouin zone.
 }
 \label{fig_srr_2}
\end{figure*}

 
\begin{figure*}[p!]
\centerline{\includegraphics[width=14cm]{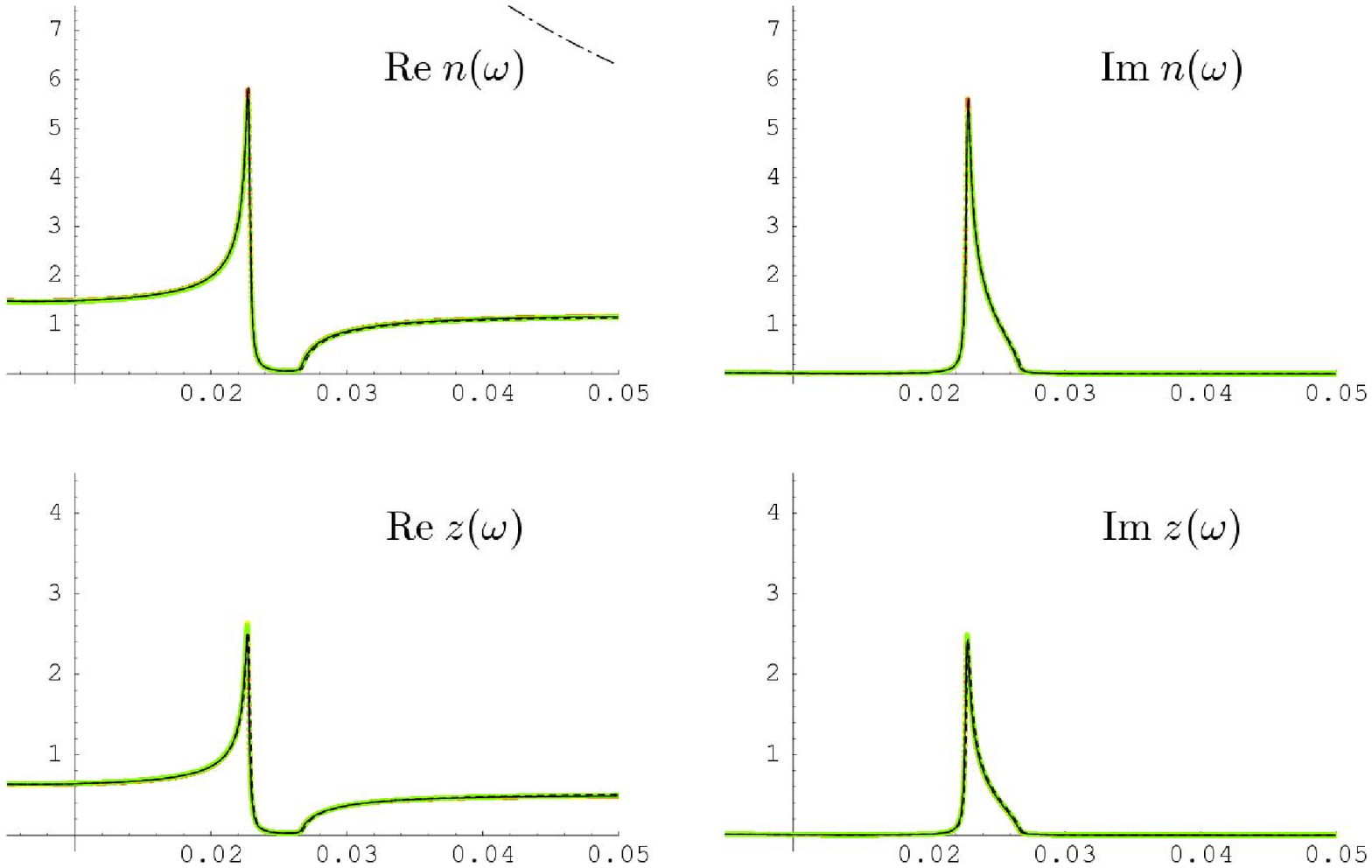}}
 \caption{%
 (color online) 
 For the simulated low-frequency SRR meta-material the effective index of refraction 
 $n_\mathrm{eff}(\omega)$ and impedance $z_\mathrm{eff}(\omega)$ are shown.  
 The colored curves represent the HEM approximation
 (Eqs.~\ref{AD_inv_scattering_z}, \ref{AD_inv_scattering_n})
 of the simulation data for the first 3 unit cells,
 the solid black line the HEM(PEM) approximation,
 and the dashed line the PEM approximation (cf.~Eqs.~\ref{R_mu_pem}, \ref{R_eps_pem})
 obtained from the first unit cell data.
 Note that far away from the edge of the Brillouin zone HEM and PEM approximation,
 and the expected homogeneous medium behavior coincide.
 }
 \label{fig_srr_low_1}
\end{figure*}

\begin{figure*}
\centerline{\includegraphics[width=14cm]{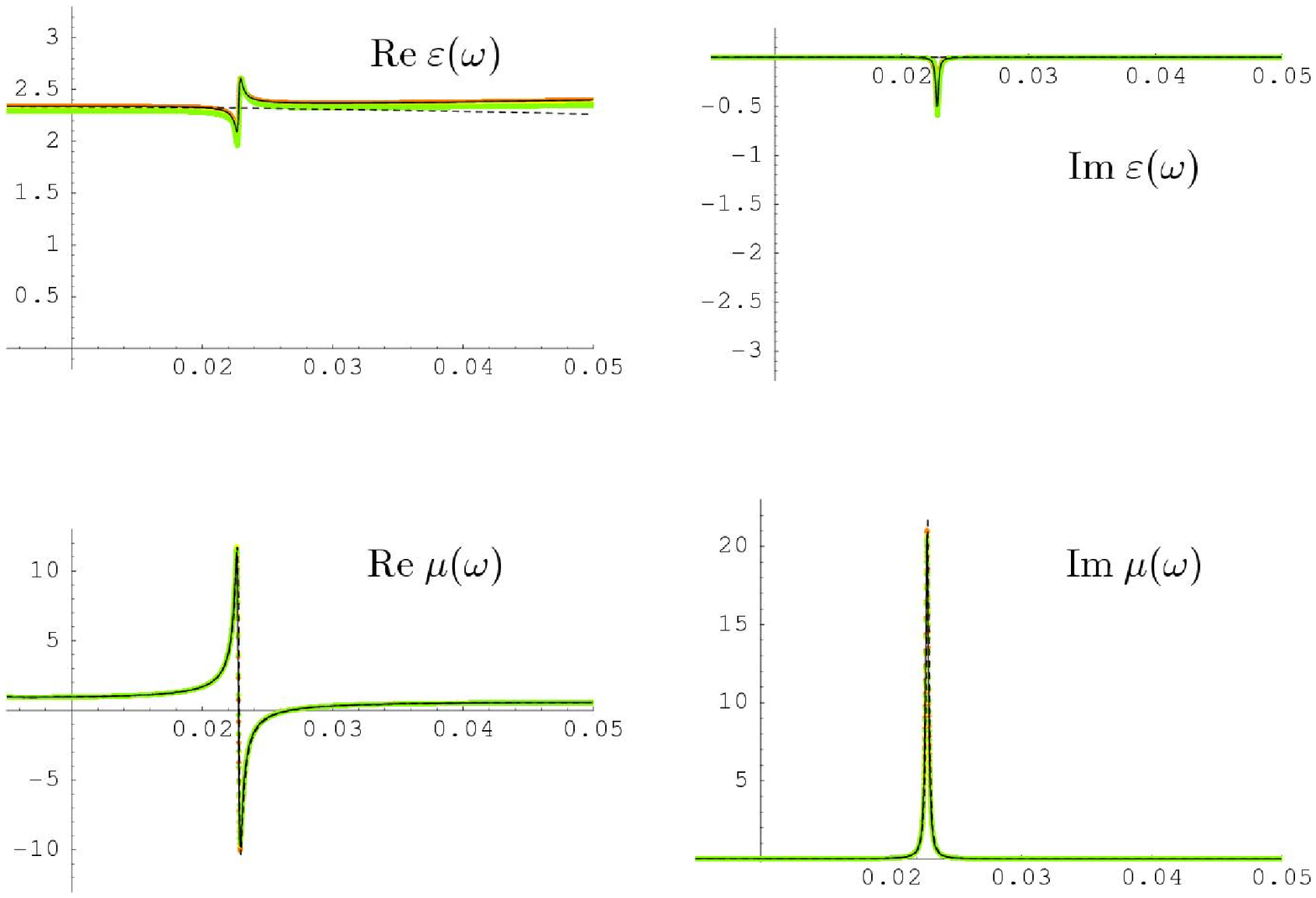}}
 \caption{%
 (color online) 
 For the simulated low-frequency SRR meta-material the effective permittivity
 $\varepsilon_\mathrm{eff}(\omega)$ and permeability $\mu_\mathrm{eff}(\omega)$
 are shown.  
 The colored curves represent the HEM approximation
 (cf.~Eqs.~\ref{AD_inv_scattering_z}, \ref{AD_inv_scattering_n})
 of the simulation data for the first 3 unit cells,
 the solid black line the HEM(PEM) approximation,
 and the dashed line the PEM approximation (Eqs.~\ref{R_mu_pem}, \ref{R_eps_pem})
 obtained from the first unit cell data.
 In the low-frequency limit the resonance/anti-resonance coupling as well 
 as the negative imaginary parts disappear.
 }
 \label{fig_srr_low_2}
\end{figure*}


\begin{figure*}[p!]
\centerline{\includegraphics[width=14cm]{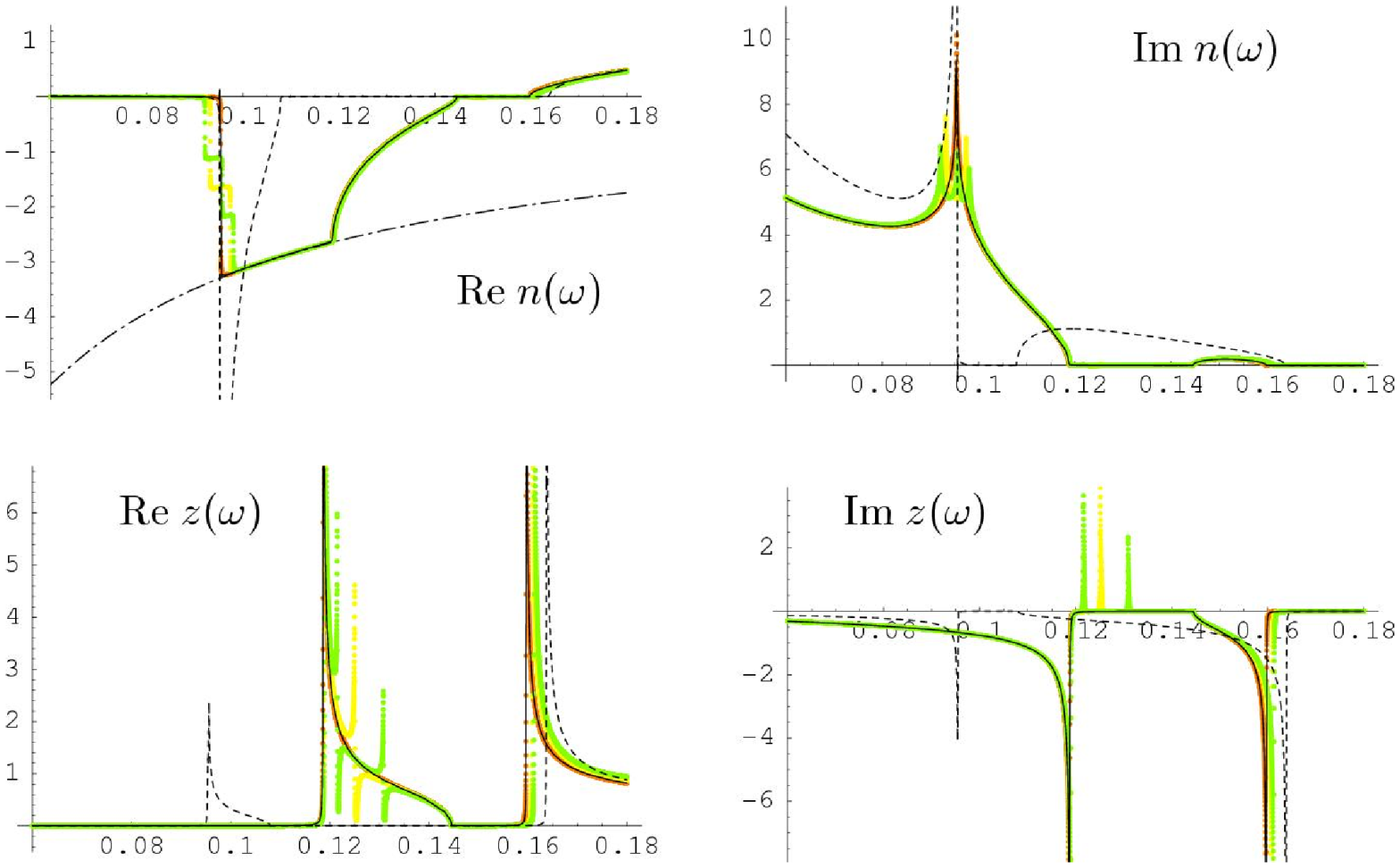}}
 \caption{%
 (color online) 
 For the simulated off-plane LHM meta-material the effective index of refraction 
 $n_\mathrm{eff}(\omega)$ and impedance $z_\mathrm{eff}(\omega)$ are shown.  
 The colored curves represent the HEM approximation
 (Eqs.~\ref{AD_inv_scattering_z}, \ref{AD_inv_scattering_n})
 of the simulation data for the first 3 unit cells,
 the solid black line the HEM(PEM) approximation,
 and the dashed line the PEM approximation (cf.~Eqs.~\ref{R_mu_pem}, \ref{R_eps_pem})
 obtained from the first unit cell data.
 }
 \label{fig_lhm_1}
\end{figure*}

\begin{figure*}
\centerline{\includegraphics[width=14cm]{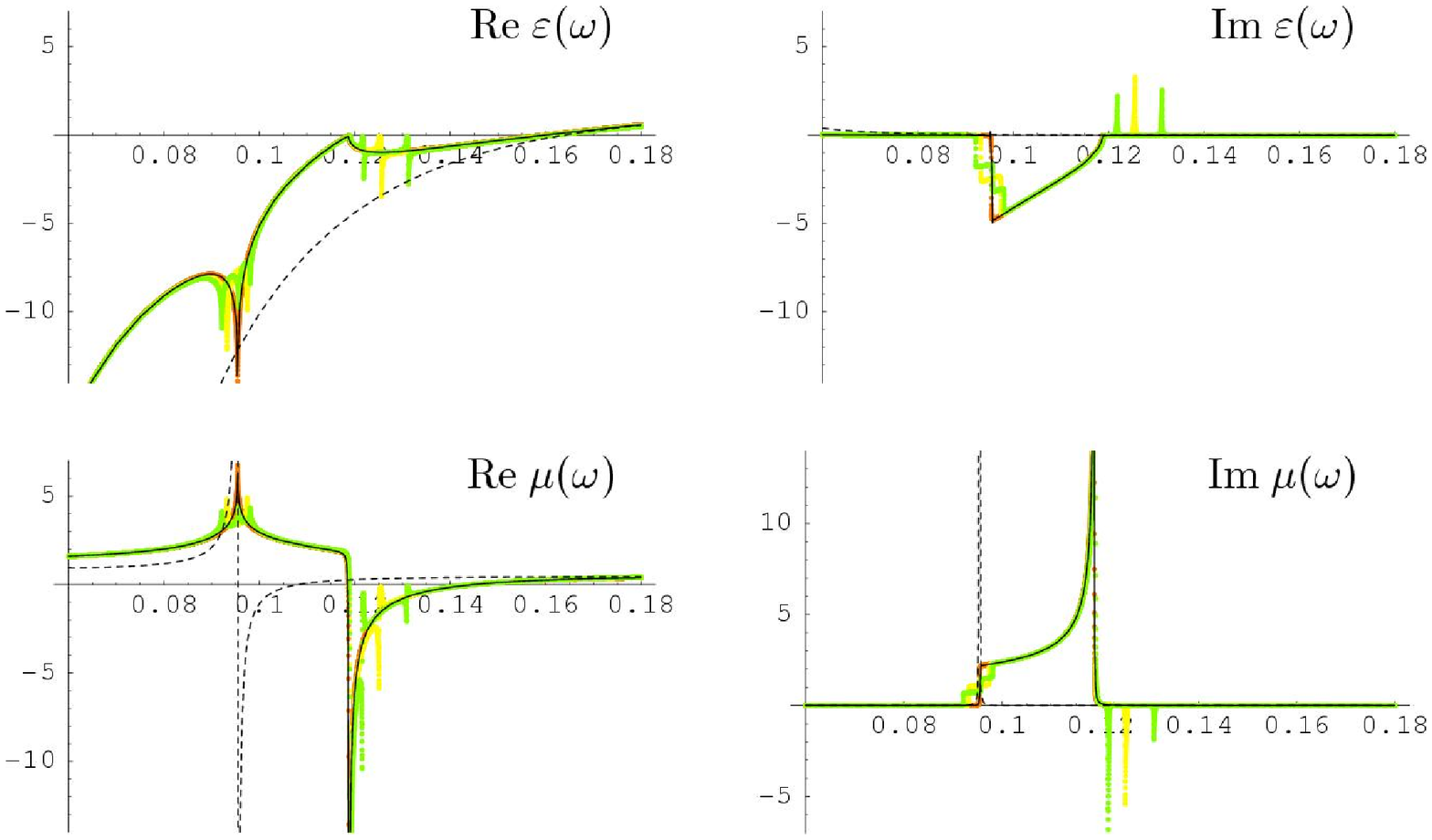}}
 \caption{%
 (color online) 
 For the simulated off-plane LHM meta-material the effective permittivity
 $\varepsilon_\mathrm{eff}(\omega)$ and permeability $\mu_\mathrm{eff}(\omega)$
 are shown.  
 The colored curves represent the HEM approximation
 (cf.~Eqs.~\ref{AD_inv_scattering_z}, \ref{AD_inv_scattering_n})
 of the simulation data for the first 3 unit cells,
 the solid black line the HEM(PEM) approximation,
 and the dashed line the PEM approximation (Eqs.~\ref{R_mu_pem}, \ref{R_eps_pem})
 obtained from the first unit cell data.
 }
 \label{fig_lhm_2}
\end{figure*}

\subsection{\label{section:results:srr} SRR}

From the naive effective medium picture we expect the SRR to expose at the
magnetic resonance brought about by the LC-oscillator type response of the 
split ring to an perpendicular external magnetic field a Lorentz-type 
resonant form\cite{Pendry99} in the permeability $\mu_\mathrm{eff}(\omega)$
but a unaffected, weakly frequency dependent permittivity 
$\varepsilon_\mathrm{eff}(\omega)$. 
If the magnetic resonance is strong enough, ie.~the imaginary part in $\mu$
is small, we should find a isolated region where 
$\mathrm{Re}\ \mu_\mathrm{eff}(\omega)<0$.
The HEM approximation of the actual simulation data is shown as effective 
impedance $z_\mathrm{hem}(\omega)$ and index of refraction 
$n_\mathrm{hem}(\omega)$ in Fig.~\ref{fig_srr_1} as the colored points.
The different colors (orange, yellow, green) correspond to different lengths
of the meta-material of one, two and three unit cells in propagation direction.
We see the typical behavior around $\omega_m$ we are already familiar with 
from previous work\cite{Koschny03b}. 
Instead of the expected form, we obtain a sharp cut-off in 
$\mathrm{Re}\ n_\mathrm{hem}(\omega)$ on the low frequency side, accompanied
by a non-zero $\mathrm{Im}\ n_\mathrm{hem}(\omega)$. 
The adjacent region with $\mathrm{Re}\ n_\mathrm{hem}\approx 0$ and 
significant $\mathrm{Im}\ n_\mathrm{hem}>0$ should correspond to the 
negative $\mu$ produced by the magnetic resonance.
Even more disturbing is the fact that the peaks in $z_\mathrm{hem}(\omega)$
which should coincide with the peaks in $n_\mathrm{hem}(\omega)$ directly 
at the resonance frequency $\omega_m$ do appear at substantially lower 
frequency.
This behavior makes it hard to determine $\omega_m$ for the meta-material 
since the retrieved $n_\mathrm{hem}(\omega)$ and $z_\mathrm{hem}(\omega)$ 
mutually disagree upon the value.
For the first time we show that the HEM approximation of the meta-material is,
apart from some additional noise for longer systems, indeed length independent.
Although we only show data for the first three unit cells we confirmed the
length independence for all system lengths up to 11 unit cells.
The dash-dotted line in Fig.~\ref{fig_srr_1}($a$) indicates the upper edge of 
the first Brillouin zone, $n_\mathrm{edge}=k_\mathrm{edge}/k=\pi/(k L)$ where
$L$ is the length of the unit cell in propagation direction.
Clearly the cut-off of $\mathrm{Re}\ n_\mathrm{hem}(\omega)$ coincides with
this line. Note also that the peaks in $z_\mathrm{hem}(\omega)$ appear exactly
when $\mathrm{Re}\ n_\mathrm{hem}(\omega)$ reaches $n_\mathrm{edge}(\omega)$. 
This behavior is generic, qualitatively the same is observed for different
sizes of the unit cell and different geometries of the SRR,
including single-ring and double-ring SRRs as well as more symmetric 
multi-gap SRRs\cite{OBrien04}.
The corresponding effective permittivity and permeability of the HEM 
approximation are shown in Fig.~\ref{fig_srr_2}. Again, the colored points
represent the simulation data for one to three unit cells.
The most striking deviation from the expected effective medium behavior
is the resonance/anti-resonance coupling between 
$\mathrm{Re}\ \mu_\mathrm{hem}(\omega)$ and 
$\mathrm{Re}\ \varepsilon_\mathrm{hem}(\omega)$, accompanied by a significant
negative imaginary part of the permittivity,
$\mathrm{Im}\ \varepsilon_\mathrm{hem}(\omega)<0$.
Moreover, the negative region of the $\mu$ resonance is strongly but 
characteristically deformed and not ascending monotonically from a negative
divergency.
Of course, the divergencies of the effective parameters would be blurred in 
the presence of large imaginary parts in $\mu$ or $\varepsilon$, but at least
for our simulations using almost perfect metals in vacuum we would expect
reasonably sharp divergencies.
As a consequence of the length independence of $n_\mathrm{hem}$ and 
$z_\mathrm{hem}$ also the retrieved $\mathrm{Re}\ \mu_\mathrm{hem}(\omega)$ 
and $\mathrm{Re}\ \varepsilon_\mathrm{hem}(\omega)$ are basically 
length independent.

In the analytic sections above we demonstrated that the periodicity can 
produce all that kind of effects violating the effective medium picture in
our simulations. Now we show that the PEM approximation of the simulation 
data yields reasonable effective parameters free of the above artifacts.
The dashed curves in Figs.~\ref{fig_srr_1} and \ref{fig_srr_2} represent 
the effective parameters
$\mu_\mathrm{pem}(\omega)$ and $\varepsilon_\mathrm{pem}(\omega)$,
with the corresponding $n_\mathrm{pem}(\omega)$ and $z_\mathrm{pem}(\omega)$, 
extracted from the PEM approximation of the simulation data as described above.
For the PEM unit cell we chose $n_a=5$, $n_d=1$ and $n_b=4$ to fit the
symmetry of the meta-materials unit cell that embeds the 1x7x7 SRR into a
6x10x10 unit cell.
From the corresponding 
$\mu_\mathrm{core}(\omega)$ and $\varepsilon_\mathrm{core}(\omega)$
we can in turn calculate the scattering amplitudes of the PEM and subject
those to the HEM inversion. 
The resulting HEM(PEM) parameters are shown as the solid black lines which
virtually coincide with the first unit cells simulated HEM parameters, 
hence basically also with those of the the longer systems,
proving that the PEM approximation is reliable. 
Obviously the PEM parameters behave exactly as we expected from the effective
medium picture in the first place. 
There is a clean resonance in $n_\mathrm{pem}(\omega)$ and the peaks of
$n_\mathrm{pem}(\omega)$ and $z_\mathrm{pem}(\omega)$ now appear at the same 
frequency, rendering $\omega_m$ well-defined.
There is no resonance/anti-resonance coupling between 
$\mu_\mathrm{pem}(\omega)$ and $\varepsilon_\mathrm{pem}(\omega)$ anymore.
The permittivity is roughly constant across $\omega_m$ and does not show
any negative imaginary part. The permeability $\mu_\mathrm{pem}(\omega)$
exposes a clear, anti-symmetric resonance at $\omega_m$ in its real part 
in conjunction with a symmetric absorption peak in the imaginary part 
at the same frequency.
Note further that the frequency $\omega_{mp}$ where 
$\mathrm{Re}\ \mu_\mathrm{hem}(\omega)$ get back to positive values after 
the magnetic resonance is substantially smaller than the
corresponding frequency for $\mathrm{Re}\ \mu_\mathrm{pem}(\omega)$.
The procedure yields analogue results for different SRR geometries and different
size of the unit cell (not shown).

Effective medium behavior was expected in the first place because the vacuum
wavelength of the incident radiation at the magnetic resonance frequency 
$\omega_m$ is large compared to the size of the unit cell for the customary 
LH and related meta-materials. 
In most experiments and simulations this ratio is in the range of 
ten to five\cite{%
Shelby01a,Shelby01b,Bayindir02,Markos02b,Smith00,Houck03,%
Koschny03b,Li03,Ozbay03,Parazzoli03,OBrien04,Yen04}. 
In the SRR meta-material presented above we went intensionally to the lower
extreme to make the effects of the periodicity more pronounced and more easy
to identify.
In Figs.~\ref{fig_srr_low_1} and \ref{fig_srr_low_2} we present the analog
HEM and PEM approximation results for a SRR with five times lower $\omega_m$.
The size of the SRR and the unit cell are kept constant.
To lower the resonance frequency of a SRR without changing the size one would
usually decrease the width of the gap in the SRR increasing its capacitance.
Due to the limitations of the used TMM implementation (uniform discretization)
this was not feasible. Therefore we adopted the alternative possibility
to place some high dielectric constant material inside the SRR gap, 
which serves the same purpose and can be used to emulate a narrower gap.
For this low-frequency SRR the wavelength to unit cell size ratio around
$\omega_m$ is roughly 25, higher than in any published LHM. 
The behavior of the retrieved effective 
$n_\mathrm{hem}(\omega)$ and $z_\mathrm{hem}(\omega)$ 
in Fig.~\ref{fig_srr_low_1} is now qualitatively as expected from the 
effective medium picture, the refractive index and impedance of the HEM 
approximation virtually coincide with the PEM parameters 
$n_\mathrm{pem}(\omega)$ and $z_\mathrm{pem}(\omega)$.
As the magnetic resonance is now shifted far below the edge of the Brillouin 
zone there is no visible cut-off in $\mathrm{Re}\ n_\mathrm{hem}(\omega)$ and 
the resonance peaks in $n$ and $z$ appear at the same frequency $\omega_m$.
Note that $\mu_\mathrm{hem}(\omega)$ reaches unity away from $\omega_m$ 
to either side.
Also the the effective permeabilities of HEM and PEM in 
Fig.~\ref{fig_srr_low_2} do coincide, exposing a clean anti-symmetric 
resonance in $\mathrm{Re}\ \mu$ and a symmetric positive absorption peak
in $\mathrm{Im}\ \mu$.
Surprisingly, though weak there is still a noticeable residue of the 
resonance/anti-resonance coupling left in 
$\mathrm{Re}\ \varepsilon_\mathrm{hem}(\omega)$ together with the 
corresponding negative imaginary part.
Only here the HEM approximation deviates from the PEM approximation which 
expectedly does not show these effects.
All effective parameters are almost perfectly length independent.
This has been verified for up to 10 unit cells in propagation direction 
(not shown).
Note also the absence of the additional noise observed in the effective
parameters for longer systems.
These results show that the artifacts in the HEM approximation which we 
identified as effects of the meta-material's periodicity vanish if we approach
the effective medium limit. 
At low frequencies HEM and PEM approximation converge, however, even in this 
extreme low-frequency limit remains of the periodicity artifacts are still
visible. 
Since for virtually all the meta-materials measured or simulated that have
been published the wavelength to unit cell size ratio is worse than 25, 
we expect the meta-material's periodicity required to be taken into account.

\subsection{\label{section:results:lhm} LHM}

The most simple way to obtain a left-handed material is to add an 
appropriately dimensioned\cite{Koschny04a,Katsarakis04b} 
continuous wire to the SRR considered above.
From the naive effective medium picture we expect for the LHM a Lorentz-type
resonance in $\mu$ in combination with a plasmonic form, modified by the 
electric cut-wire response of the SRR where necessary\cite{Koschny04a}, 
which is essentially negative around the magnetic resonance frequency 
$\omega_m$. 
The effective impedance and index of refraction of the HEM approximation for 
the first three unit cells of the LHM in propagation direction is shown in
Fig.~\ref{fig_lhm_1} as the colored points.
In violation of the assumed effective medium picture we find again a cut-off
of the negative resonant $\mathrm{Re}\ n_\mathrm{hem}(\omega)$, this time 
at the lower edge of the first Brillouin zone,
$n_\mathrm{edge}=-k_\mathrm{edge}/k=-\pi/(k L)$ where
$L$ is the length of the unit cell in propagation direction.
The imaginary part of $n$ does not have the expected form either.
In $z_\mathrm{hem}(\omega)$ we expect two peaks for the LHM, one at $\omega_m$
and another one at the electric (effective) plasma frequency $\omega_p'$
which is the lowest frequency where $\varepsilon(\omega)$ becomes positive.
Though the retrieved $z_\mathrm{hem}(\omega)$ does show two such peaks, the
position of the first one associated with $\omega_m$ does not agree with
the $\omega_m$ derived from $n_\mathrm{hem}(\omega)$. 
This is the same issue as found for the SRR above.
The corresponding effective permeability $\mu_\mathrm{hem}(\omega)$ and 
permittivity $\varepsilon_\mathrm{hem}(\omega)$ 
are shown in Fig.~\ref{fig_lhm_2}.
The previously published resonance/anti-resonance coupling in the real parts 
around $\omega_m$ is clearly visible, together with the appearance of the
negative imaginary part in the permittivity and the misshapen absorption peak
in the the permeability.
Also for the LHM we can now confirm the length independence of the HEM 
approximation, apart from some additional noise, up to 10 unit cells in 
propagation direction. 
As for the SRR the PEM approximation of the LHM yields effective parameters 
free of all the artifacts seen in the HEM parameter which demonstrates again
their origin in the periodic structure of the meta-material.
In $n_\mathrm{pem}(\omega)$ and $z_\mathrm{pem}(\omega)$ we obtain an
untruncated magnetic resonance and agreement of $n$ and $z$ upon the position
of $\omega_m$.
The permittivity $\varepsilon_\mathrm{pem}(\omega)$ and permeability
$\mu_\mathrm{pem}(\omega)$ show a clean magnetic resonance with a symmetric
absorption peak and the anticipated electric plasmonic behavior without any
negative imaginary parts. 
In $\varepsilon_\mathrm{pem}(\omega)$ we can even recognize the beginning
ascent of the imaginary part to the absorption peak at $\omega=0$ contributed
by the plasma resonance of the continuous wire.
Note again the shift in $\omega_{mp}$ while the electric plasma frequency
$\omega_p'$ is essentially the same for HEM and PEM approximation, 
although the descent to negative values for decreasing frequency is more 
rapid for $\varepsilon_\mathrm{pem}(\omega)$.
Also for the LHM these results are generic, ie.~have been qualitatively 
reproduced for different sizes of the SRR and continuous wire components
of the LHM and the unit cell.


\begin{figure*}[p!]
\centerline{\includegraphics[width=14cm]{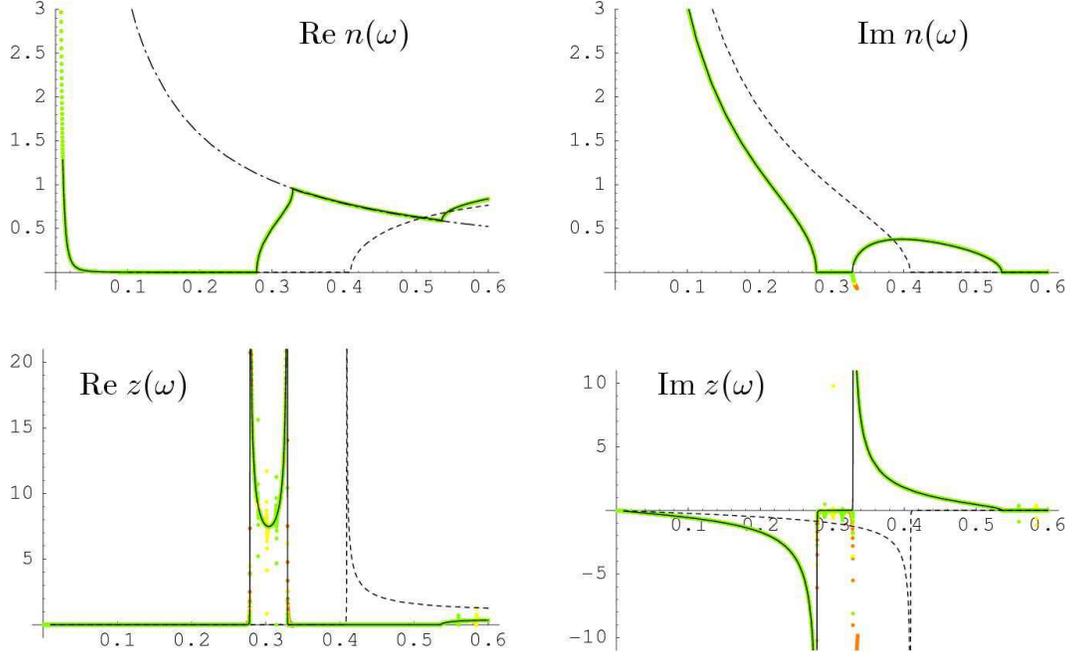}}
 \caption{%
 (color online) 
 For the simulated continuous wire meta-material the effective index of refraction 
 $n_\mathrm{eff}(\omega)$ and impedance $z_\mathrm{eff}(\omega)$ are shown.  
 The colored curves represent the HEM approximation
 (Eqs.~\ref{AD_inv_scattering_z}, \ref{AD_inv_scattering_n})
 of the simulation data for the first 3 unit cells,
 the solid black line the HEM(PEM) approximation,
 and the dashed line the PEM approximation (cf.~Eqs.~\ref{R_mu_pem}, \ref{R_eps_pem})
 obtained from the first unit cell data.
 }
 \label{fig_wire_1}
\end{figure*}

\begin{figure*}
\centerline{\includegraphics[width=14cm]{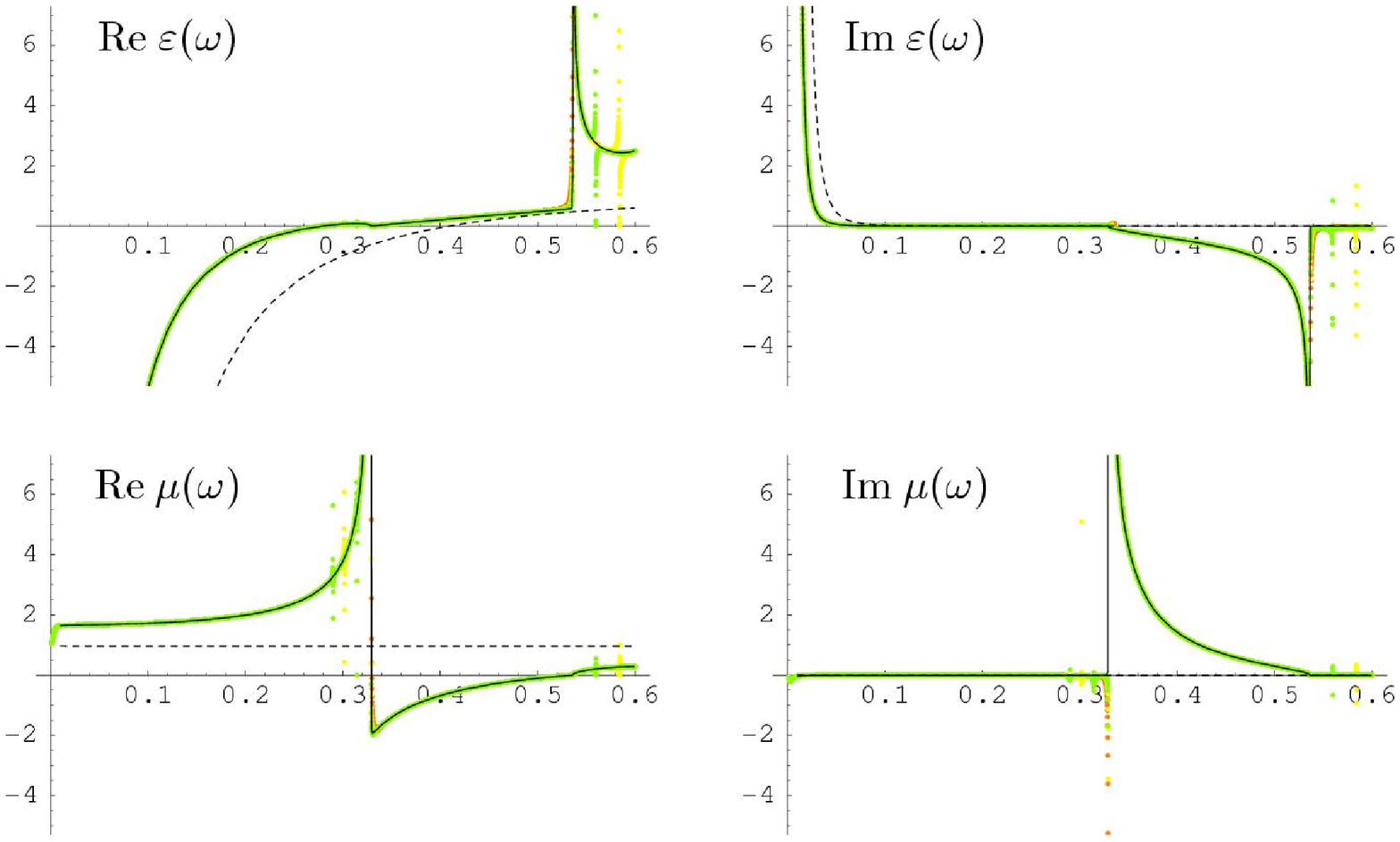}}
 \caption{%
 (color online) 
 For the simulated continuous wire meta-material the effective permittivity
 $\varepsilon_\mathrm{eff}(\omega)$ and permeability $\mu_\mathrm{eff}(\omega)$
 are shown.  
 The colored curves represent the HEM approximation
 (cf.~Eqs.~\ref{AD_inv_scattering_z}, \ref{AD_inv_scattering_n})
 of the simulation data for the first 3 unit cells,
 the solid black line the HEM(PEM) approximation,
 and the dashed line the PEM approximation (Eqs.~\ref{R_mu_pem}, \ref{R_eps_pem})
 obtained from the first unit cell data.
 }
 \label{fig_wire_2}
\end{figure*}


\begin{figure*}[p!]
\centerline{\includegraphics[width=14cm]{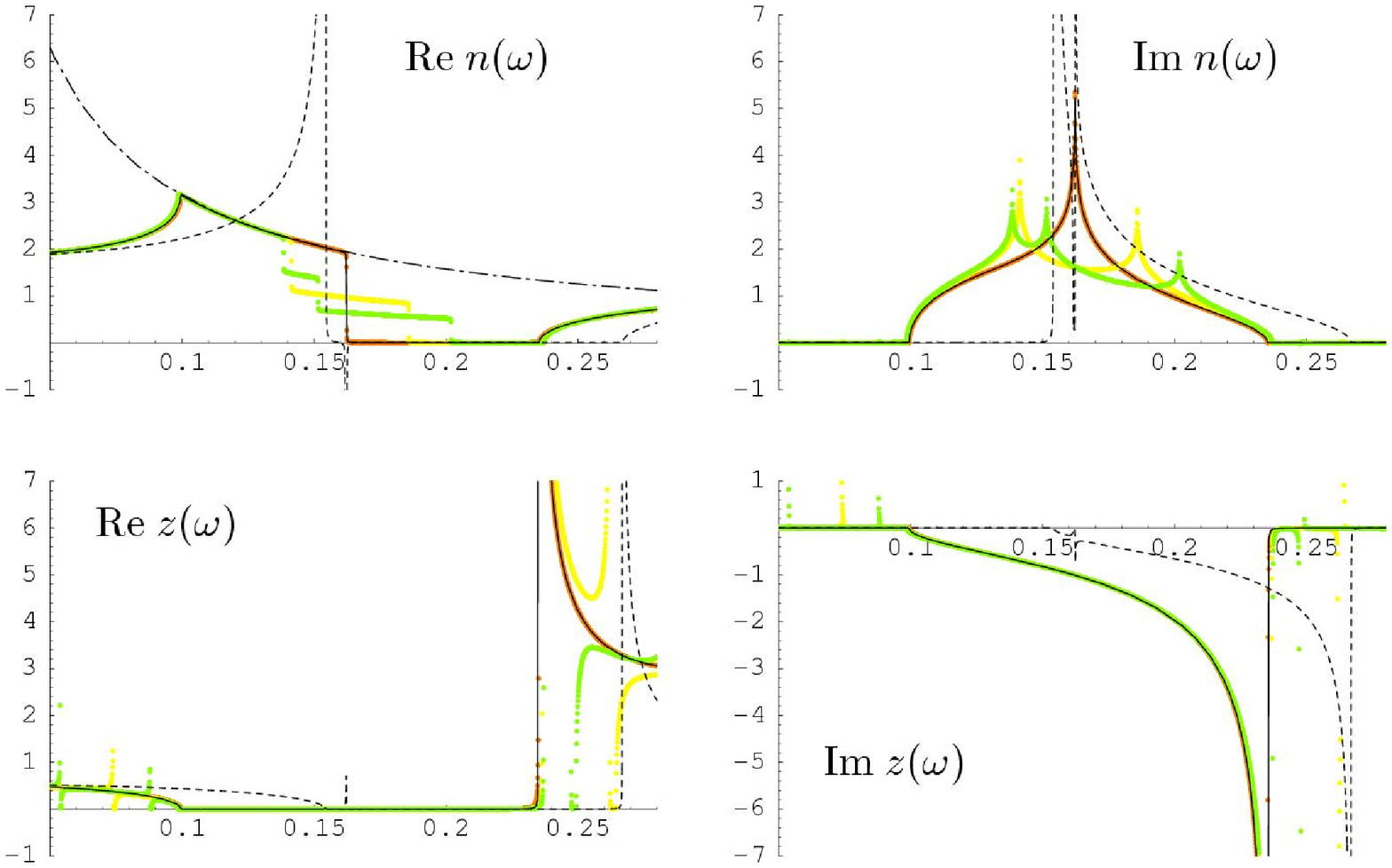}}
 \caption{%
 (color online) 
 For the simulated cut-wire meta-material the effective index of refraction 
 $n_\mathrm{eff}(\omega)$ and impedance $z_\mathrm{eff}(\omega)$ are shown.  
 The colored curves represent the HEM approximation
 (Eqs.~\ref{AD_inv_scattering_z}, \ref{AD_inv_scattering_n})
 of the simulation data for the first 3 unit cells,
 the solid black line the HEM(PEM) approximation,
 and the dashed line the PEM approximation (cf.~Eqs.~\ref{R_mu_pem}, \ref{R_eps_pem})
 obtained from the first unit cell data.
 The weak additional structure in the PEM approximation close to the resonance
 is indicating the beginning breakdown of the approximation of the meta-material
 within our most simple periodic medium model.
 }
 \label{fig_cutwire_1}
\end{figure*}

\begin{figure*}
\centerline{\includegraphics[width=14cm]{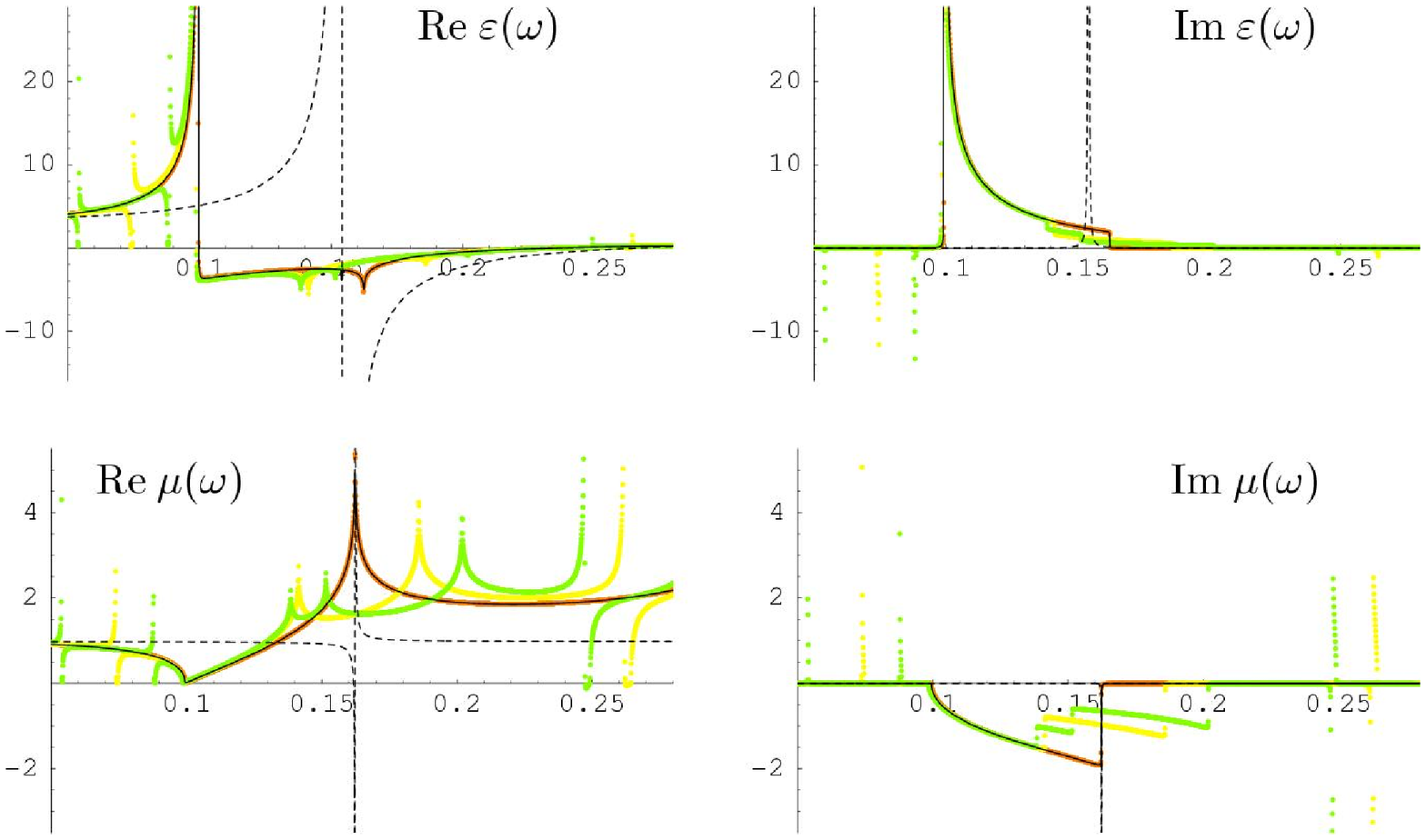}}
 \caption{%
 (color online) 
 For the simulated cut-wire meta-material the effective permittivity
 $\varepsilon_\mathrm{eff}(\omega)$ and permeability $\mu_\mathrm{eff}(\omega)$
 are shown.  
 The colored curves represent the HEM approximation
 (cf.~Eqs.~\ref{AD_inv_scattering_z}, \ref{AD_inv_scattering_n})
 of the simulation data for the first 3 unit cells,
 the solid black line the HEM(PEM) approximation,
 and the dashed line the PEM approximation (Eqs.~\ref{R_mu_pem}, \ref{R_eps_pem})
 obtained from the first unit cell data.
 }
 \label{fig_cutwire_2}
\end{figure*}

\subsection{\label{section:results:wcw} Continuous wire and cut-wire}

Not only the behavior of the SRR based meta-materials around the magnetic 
resonance but also that of the periodic arrangement of continuous wires or 
cut-wires are strongly affected by the periodicity. 
In the effective medium picture, the continuous wire is expected to expose  
a simple plasmonic permittivity going monotonously from negative to positive 
real part and crossing zero at a single plasma frequency $\omega_p$.
In the HEM approximation of simulated continuous wire 
meta-materials\cite{Pendry96b}, 
eg.~the isolated continuous wire array from 
the LHM discussed above,
we observe the anticipated plasmonic behavior only at low frequencies up to
the order of the plasma frequency $\omega_p$ of the wire. 
At higher frequency we find a sequence of additional stopbands 
which can be explained as periodicity band gaps in the framework of the 
PEM model: 
whenever the residue class $\mathrm{Re}\ n(\omega)\ \mathrm{mod}\ 2\pi/k$
comes close to the edge of the first Brillouin zone a periodicity band gap
is opened where a representative of $\mathrm{Re}\ n_\mathrm{hem}(\omega)$ 
follows the line
$n_\mathrm{edge}(\omega)$ and $\mathrm{Im}\ n_\mathrm{hem}(\omega)$ 
is significantly non-zero.
At the boundaries of each of these band gaps the real and the imaginary part
of the effective impedance $z_\mathrm{hem}(\omega)$ have either a zero or 
a pole which leads to the appearance of a alternating sequence of phony
resonance-like structures in 
$\mu_\mathrm{hem}(\omega)$ and $\varepsilon_\mathrm{hem}(\omega)$.
The corresponding series of transmission above the ``first'' $\omega_p$
have also been confirmed in experiments\cite{nikos-pc-03} 
with thin metallic wires on PCB boards.
Using the PEM approximation of the simulation data we can describe the 
scattering amplitudes in terms of a $\varepsilon_\mathrm{pem}(\omega)$ 
which does possess just the expected plasmonic form in conjunction with 
an almost exactly $\mu_\mathrm{pem}(\omega)=1$. 
The plasma frequency in $\varepsilon_\mathrm{hem}(\omega)$ however does not 
coincide with the ``lowest'' $\omega_p$ of the HEM approximation but appears
moderately shifted to higher frequency.
The HEM and PEM approximation for a meta-material comprising a periodic
array of continuous thin wires parallel to the electric field of the
incident electromagnetic wave is shown in Figs.~\ref{fig_wire_1} and 
\ref{fig_wire_2}.
Both effective material approximations are virtually independent on the 
system length.

The periodic array of cut-wires is of interest for two reasons: first,
it can be used as a model of the electric response of the SRR and second,
we could substitute the continuous wire in the LHM to tailor the collective
$\varepsilon_\mathrm{eff}(\omega)$ or to simplify its mechanical construction.
For a sufficiently low resonance frequency $\omega_e$ the cut-wire shows a
behavior analogue to the SRR but with the electric and magnetic parameters
exchanged. 
We observe\cite{Koschny03b} all the previously described artifacts in the 
HEM approximation: the resonance/anti-resonance coupling, where the (electric)
resonance appears this time in $\varepsilon_\mathrm{hem}(\omega)$, accompanied
by an negative imaginary part in $\mu_\mathrm{hem}(\omega)$, 
the cut-off of $n_\mathrm{hem}(\omega)$ at the Brillouin zone edge, and so on.
However, since in real meta-materials the cut-wire resonance usually 
appears at much higher frequency, we observe an additional periodicity 
band gap at lower frequency, well separated from the cut-wire response, 
qualitatively as shown in Fig.~\ref{fig_ana_srr}. 
The effective parameters $\mu_\mathrm{hem}(\omega)$ and 
$\varepsilon_\mathrm{hem}(\omega)$ at the lower boundary of this additional 
band gap will appear very similar to those of a cut-wire resonance at low 
frequency $\omega_e$.
Note that we can shift the cut-wire resonance to arbitrarily low frequency
by reducing the gap in the longitudinal direction of the cut-wires, 
ie.~making the finite wires almost as long as the width of the unit cell.
The HEM and PEM approximation for such a meta-material of cut-wires 
parallel to the electric field of the incident electromagnetic wave 
is shown in Figs.~\ref{fig_cutwire_1} and 
\ref{fig_cutwire_2}.
In either case, the effective parameters obtained from the PEM approximation
do basically not show any artifacts around the cutwire resonance nor the 
periodicity band gaps and behave very much like expected from the naive
effective medium picture.
This gives us a criterion to distinguish the real cut-wire resonance in a
meta-material from the possible phony one brought about by the periodicity.

\section{\label{section:discussion} Discussion}

When Veselago conceived the idea of heft-handed electromagnetic material
he essentially considered theoretical homogeneous media for which there are
no correspondents in nature.
Though there are materials exposing a negative magnetic response and other
materials with a negative electric response, the challenge is to obtain both
simultaneously in a material and, moreover, at an experimentally useful
frequency in or above the microwave range.
After Pendry proposed the first practical possibility to obtain controllable
resonant magnetic and plasmonic negative electric response by the means of
the geometric shape of electric conductors (SRR and continuous wire)
in a periodic arrangement,   
numerical simulations have been conducted, attempting to verify the 
simultaneous negative magnetic and electric response.
Note that because of the technical difficulty to obtain reliable complex 
reflection amplitudes from experimental measurements numerical simulations
are here the most important tool to address the question of negative 
electromagnetic response.
The existence of negative refraction, which is merely a consequence of an
anti-parallel phase and group velocity, has been demonstrated experimentally
but provides no prove for the actual Veselago picture of simultaneously
negative $\mu$ and $\varepsilon$.
For this purpose, at typical vacuum wavelengths of around ten times the size 
of the unit cell effective medium behavior of the meta-materials has been 
assumed throughout the literature such that an effective permeability and 
permittivity could be obtained in a HEM approximation.
Although the working frequency of a SRR-based LHM is theoretically arbitrary,
the fabrication technique in experiments and the limited computer power in
numerical simulations impose certain constraints on the size of the 
smallest structures (particularly the width of the gap in the SRR) in
comparison to the size of the unit cell. 
As a consequence the ratio of the vacuum wavelength around the magnetic 
resonance to the size of the unit cell is confined as well.

\medskip
\paragraph*{Continuum HEM approximation.}
The effective refractive index $n_\mathrm{hem}(\omega)$, 
impedance $z_\mathrm{hem}(\omega)$ and, derived from those, 
permeability $\mu_\mathrm{hem}(\omega)$ and
permittivity $\varepsilon_\mathrm{hem}(\omega)$ published by 
us\cite{Markos02b,Smith02b,Markos03a,Koschny03b} and 
others\cite{Parazzoli03,OBrien02,Chen04,OBrien04}
for the LHM, SRR and cut-wire meta-materials do only in first approximation 
meet the the anticipated effective response of the material. 
For the considered polarization (electric field parallel to the continuous 
wire and perpendicular to the gap bearing side of the SRR) and direction of 
propagation in the SRR plane the resonant circulating currents inside the
SRR ring should couple and respond only to the magnetic field, 
affecting the behavior of the effective $\mu(\omega)$.
The electric field couples to the continuous wire or cut-wire. 
However, it also couples to a separate electric resonance of the SRR 
caused by induced polarization currents oscillating linearly in the gap-less
sides of the SRR which are parallel to the electric field. 
For other orientations, due to the anisotropy arising from the gap in the SRR 
ring, the electric field may also couple to the resonance of the circulating
currents making the total effective behavior a lot more complicated.
In some cases violating the inversion symmetry of the unit cell also a 
magnetic coupling to the electric cut-wire response is possible.
Though we obtain from the simulations a resonant magnetic 
(magnetic resonance frequency $\omega_m$ for SRR, LHM) 
and electric (cut-wire resonance frequency $\omega_e$ for SRR, cut-wire) 
response as well as basically the plasmonic response of the continuous wire 
and its corresponding contribution in the LHM, 
the correspondence to the effective medium picture is spoiled by
partially very significant anomalies: 
($i$) Resonance/anti-resonance coupling. 
We expect the electric and magnetic response of the discussed 
meta-materials to be independent. However, whenever there is a resonance in 
$\mathrm{Re}\ \mu$ we simultaneously observe an anti-resonant behavior in 
$\mathrm{Re}\ \varepsilon$, and vice versa. 
The anti-resonant structures in the real part are accompanied by an negative
imaginary part.
($ii$) Misshapen, truncated resonances. 
The divergence in $\mathrm{Re}\ n$ appears to be cut-off at the edges of the 
first Brillouin zone and, in particular, the negative regions of the magnetic
resonance in $\mu$ and cut-wire resonance in $\varepsilon$ do not return from
large negative real part but seem to saturate in a rather shallow behavior.
The corresponding absorption peak in the imaginary parts as misshapen and 
highly asymmetric too.
($iii$) Discrepancy between $n$ and $z$ about the positions of the resonances.
We expect the peaks (or zeros) in the index of refraction and the impedance 
to appear exactly at the resonance frequencies $\omega_m$ and $\omega_e$, 
or the effective plasma frequency $\omega_p'$. 
From the simulations, however, we find different frequencies from $n$ and $z$,
respectively. 
This lead for instance to an ``internal structure'' of the magnetic resonance
as shown in Fig.~\ref{fig_srr_1}, which can not easily be explained within 
the assumed effective medium picture.
($iv$) Additional spectral structures.
Apart from structures around the anticipated contributions of the 
meta-material's constituents $\omega_m$, $\omega_e$ and $\omega_p'$ we observe
a lot of additional structure, especially at higher frequency, which can not
be accounted for. 
The observed artifacts in the HEM approximations are quite generic and have 
meanwhile been reported by different groups.
An explanation for all these effects is given by the periodicity, see below.

The occurrence of negative imaginary parts the permittivity or permeability 
has been criticized by several authors\cite{Depine04,Efros04b}. 
Indeed, a physical, passive homogeneous material may not possess any
negative imaginary in $\mu$ and $\varepsilon$. 
As long as the material is passive, it can only absorb energy from the electromagnetic field. If there was any negative imaginary part in $\mu$ or 
$\varepsilon$ one could device a geometry of this material which would violate
the passivity.
This requirement does not apply to the HEM approximation. 
The {\em effective} material is defined as the homogeneous material which
reproduces the scattering amplitudes of the meta-material within the given
geometry, ie.~normal incidence to a homogeneous slab of finite length in 
propagation direction, if a length independent solution does exist. 
Our retrieved HEM parameters provide such a length independent description
of the actual scattering amplitudes of the meta-material.
For the (stationary) scattering of plain waves at a finite homogeneous slab
in the continuum there are $\mu$ and $\varepsilon$ with negative 
imaginary parts which do not violate the passivity of the material, 
as long as much weaker conditions  
$\mathrm{Im}\,\mu + \mathrm{Im}\,\varepsilon \ge 0 %
 \ \wedge\ %
 \mathrm{Im}\,\mu/|\mu| + \mathrm{Im}\,\varepsilon/|\varepsilon| \ge 0$
are satisfied.
If we further restrict the scattering setup requiring the thickness of slab 
to be an integral multiple of the unit cell, which is reasonable for this 
type of meta-material, we have even more freedom in $\mu$ and $\varepsilon$.
The HEM approximation is only valid for a given geometry, there is not 
necessarily a physical material exposing the same material parameters in
an arbitrary setup.

Some authors\cite{Marques02} 
have suggested a more general effective description\cite{Bassiri88} of the
meta-materials which employ tensorial $\mu$ and $\varepsilon$ to take the
coupling of the electric field to, and the anisotropy of the SRR into account.
Though this is certainly an issue for arbitrary orientation of the SRR and
will gain importance in more-dimensional materials, it is not directly 
related to the deviations from the effective medium behavior discussed above.

Apart form some additional ``noise'', we could confirm the length independence
of the HEM approximation for all considered meta-materials up to 11 unit cells
in propagation direction.
In contrast to previous work this became possible by the introduction of a
explicitly isotropic material-discretization in the TMM and the use of a
carefully chosen symmetry of the unit cell to avoid the occurrence
of cross-polarization scattering amplitudes\cite{Koschny04c}.
In the presence of cross-polarization terms the second polarization can 
contribute\cite{Markos03a} to the $T$ and $R$ of the considered polarization,
eg.~the transmission $t_{22}\sim 1$ for the passive polarization of the SRR 
which has the magnetic field in-plane and the electric field parallel 
to the gap-less sides can cut-off the transmission in the active polarization
of the SRR in the stop-bands,
$t_{11}(N)\sim t_{11}^N + t_{12}\,t_{22}^{N-2}\,t_{21} + \ldots\,$,
which should decay exponentially with the system length.
This effect can cause a phony length dependence in the HEM approximation which
is diagonal in the polarizations.

An other issue we have to address is the vacuum wavelength to unit cell size
ratio. In the present paper we have: 
1 (unit cell), 5 (SRR), 25 (low frequency SRR);
a typical value for the meta-materials found in the literature is 
$\approx 10$ or worse.
Strictly speaking we can expect effective medium behavior only in the 
$\omega\to 0$ limit. As an approximation, it may hold if the wave length
inside the structure is large compares with the typical length scale of the
meta-material provided by the size of the unit cell.
Two points are important here: 
The relevant wavelength is the wavelength inside
the structure and can be much smaller than the corresponding vacuum wavelength
if we approach the resonances at $\omega_m$ and $\omega_e$ and the magnitude
of the real part of the effective index of refraction becomes large. 
Second, we cannot know a priori how large the above ratio has to be to reach 
a reasonable effective medium behavior.
Our simulations support both points as the deviations from the expected 
behavior happen essentially when the wavevector inside the structure,
$q=n(\omega)\,k$, becomes comparable with the edge of the Brillouin zone, 
ie.~the wavelength comparable to the unit cell size;
for the low frequency SRR we find pretty good effective medium behavior
without the afore mentioned anomalies apart from a small region around 
the resonance. 
The simulations indicate that to obtain a reasonable effective medium behavior
the wavelength to unit cell size ratio has to be of the order of 30.
The less losses occur in the resonances the larger the peak in $n$, and 
consequently the required wavelength to unit cell size ratio will be.

\medskip
\paragraph*{Continuum PEM model.}
The analytic HEM approximation of the periodic medium model proves that the 
observed deviations from the anticipated effective medium behavior in the 
HEM approximation of the real simulated meta-materials can be caused by 
the periodicity or, more precisely, the reduced translational symmetry to 
the discrete group generated by the unit cell.
In particular, the model shows that those artifacts are not related to the 
actual geometric definition of the SRR and LHM resonances since the 
homogeneous core in the model just uses the resonant forms of $\mu(\omega)$
and $\varepsilon(\omega)$ without any reference to their microscopic source.

The artifacts observed in the HEM approximation ultimately originate from 
the occurrence of band gaps introduced by the periodicity, analogously to,
for instance, the band-structure in a crystal. 
In our case, the only substantial difference is the explicit frequency 
dependence of the microscopic material properties, 
ie.~the resonant forms of $\mu(\omega)$ and $\varepsilon(\omega)$ 
for the homogeneous core of the periodic medium model's unit cell.
These periodicity band gaps are distinct from the intrinsic band gaps which
arise directly from the negative product $\varepsilon\mu$ in the emulated 
response of the constituents.
To illustrate the behavior in the periodicity band gaps we assume for the 
moment $\mu(\omega)$ and $\varepsilon(\omega)$ of the core to be real.
Then $n(\omega)$ and $z(\omega)$ in Eq.~\ref{AD_inv_scattering_neff_1} 
are simultaneously either real or imaginary 
and consequently the r.h.s.\@ of (\ref{AD_inv_scattering_neff_1}) 
is real as well.
A periodicity band gap occurs whenever the r.h.s.\@ grows outside the
domain $[-1,1]$ of the cosine for a real argument: 
For the principal branch\cite{ARCCOS}, 
$n_\mathrm{eff}$ acquires a non-zero imaginary part inside the gap and
$\mathrm{Re}\ n_\mathrm{eff}(\omega)$ retains the value of zero or $\pi/(kL)$ 
for all r.h.s.\@ of equation (\ref{AD_inv_scattering_neff_1}) above $1$ or
below $-1$, respectively.
Adding a small imaginary part in $\mu(\omega)$ and $\varepsilon(\omega)$ 
of the core, as should be a good approximation for the emulated meta-material
if we do not come too close to the resonances (absorption peaks), 
adds a small imaginary part to $n_\mathrm{eff}(\omega)$ and causes 
$\mathrm{Re}\ n_\mathrm{eff}(\omega)$ to deviate slightly from $0$ or 
$\pi/(kL)$ towards $\pi/(2kL)$ inside the periodicity band gap.
Having thus established the confinement of the effective index of refraction,
$\mathrm{Re}\ n_\mathrm{eff}(\omega)$, to the edge(s) of the fist Brillouin 
zone or to zero, the coupling of $\mu$ and $\varepsilon$ 
follows as a direct consequence\cite{Koschny03b}. 
In simple words: If $n^2=\varepsilon\mu$ is confined and either one of 
$\varepsilon$, $\mu$ exposes resonant behavior the other has to go to zero
simultaneously.
This also explains why the resonance/anti-resonance coupling and the
negative imaginary parts occur only across the periodicity band gaps 
associated with the resonances but not outside.
For the general case this qualitative behavior is complicated by the 
non-zero imaginary parts.

Note again that we can obtain the real part of the index of refraction 
only as a residue class 
$\mathrm{Re}\ n_\mathrm{eff}(\omega)\ \mathrm{mod}\ 2\pi/(kL)$
which becomes immediately clear either from the length-independence of the
HEM approximation of the PEM according to equations (\ref{AD_T_approx_a},
\ref{AD_T_approx_b}) or from the argument about the simultaneous congruences
for different system lengths discussed in section \ref{section:hem}.  
Therefore the pieces of $\mathrm{Re}\ n_\mathrm{hem}(\omega)$ that follow 
the multiples of the first Brillouin zone's edge just coincide with either
zero or the first Brillouin zone's edge (upper and lower are equivalent)
itself.

An other important observation is that a periodicity band gap may occur
in between the magnetic resonance $\omega_m$ and the cut-wire resonance
$\omega_e$. 
Unfortunately, the corresponding effective $\varepsilon$ 
looks like an electric resonance with the attendant anti-resonant structure 
in $\mu$. 
This may easily be mistaken as the cut-wire resonance which would be 
expected to follow the magnetic resonance as the next feature in the 
frequency spectrum. 
Only in the low-frequency limit the latter behavior is actually observed.
Not that the parameter dependence of the phony electric resonance frequency
at the lower edge of the periodicity band gap will qualitatively
resemble the behavior of the real cut-wire resonance frequency $\omega_e$.

In comparison to the TMM simulations we see, however, too much structure,
ie.~a series of periodicity band gaps instead of only one, around the
intrinsic resonances.
The basic difference between the TMM simulation and the analytic calculation
of the scattering amplitudes in the continuum is the presence of a finite
discretization mesh in the TMM, which implies a smallest distance, 
hence in turn a largest supported momentum. 
Obviously, this limitation will become visible where $n_\mathrm{eff}(\omega)$ 
grows large.

\medskip
\paragraph*{Lattice PEM approximation.}
In our TMM simulations the lattice version of the PEM does better correspond
to the numerical data than the continuum PEM. 
This does also apply to independent Microwave Studio simulations
which, in contrast to our TMM simulations, utilize a non-uniform 
discretization of the meta-material.
If the discretization mesh is chosen finer the lattice PEM gradually
approached the continuum PEM behavior.

As discussed above the HEM and the PEM approximation of the real meta-material
are basically length-independent, 
longer systems expose the same spectral features as the first unit cell. 
However, systems that consist of more than a single unit cell in propagation 
direction do contribute {\em additional} tiny resonance-like structures 
in the effective material constants
$\mu_\mathrm{hem}(\omega)$, $\varepsilon_\mathrm{hem}(\omega)$
which are also present in the effective parameters of the PEM approximation,
$\mu_\mathrm{pem}(\omega)$ and $\varepsilon_\mathrm{pem}(\omega)$ (not shown).
For a slab of $N$ unit cells in propagation direction these additional
structures appear as tiny band gaps, 
quite similar to the periodicity band gap discussed above,
at frequencies where 
$n_\mathrm{pem}(\omega)\approx m\pi/(N kL)$ 
with $m\in[-N,N]\subset\mathbb{Z}$, ie.~whenever the effective refractive 
index derived from the simple unit cell comes close to the multiples of 
the first Brillouin zone's edge for the whole slab.
The additional structures weaken and eventually ceases to be visible 
in the low-frequency limit as can be seen, for instance,
for the SRR in Figs.~\ref{fig_srr_low_1} and \ref{fig_srr_low_2}.
Note that again the behavior is generic: it appears in our TMM simulations
for LHM, SRR and also cut-wire and continuous wire meta-materials (not shown), 
it has also been verified in Microwave Studio simulations using a different
numeric technique.
As there is no such length dependence in the analytic periodic medium model,
we interpret this additional ``noise'' as a limitation of the 
PEM approximation of the real meta-material which starts to see some internal
structure apart from the explicit periodicity.
It is not yet clear whether the groups of $N$ peaks directly at the 
magnetic resonance are related to this problem.
For these there are at least two other interpretations:
They may be caused by the coupling of successive SRRs in propagation direction
which would lead to the splitting of the resonance frequency 
$\omega_m$ as for the eigenfrequency of coupled identical oscillators, 
or the finite accuracy of the numeric simulation data could, in particular 
around the resonances, lead to an residual explicit length dependence.

We demonstrate above that the PEM approximation of real LHM and SRR 
meta-materials is good in the region around $\omega_m$ 
if the corresponding vacuum wavelength in as small as five times the 
size of the unit cell. 
Then we obtained the anticipated effective behavior in 
$\mu_\mathrm{hem}(\omega)$, $\varepsilon_\mathrm{hem}(\omega)$ 
and all the anomalies and additional features seen in the HEM approximation
arose from the explicit periodicity.
If we move to even higher frequency where the vacuum wavelength is close
to the size of the unit cell, 
also $\mu_\mathrm{hem}(\omega)$ and $\varepsilon_\mathrm{hem}(\omega)$ 
start to develop unexpected features like additional magnetic response
around the cut-wire resonance.
Although even then the PEM spectrum is not nearly as erratic as the 
corresponding HEM spectrum, this indicates the breakdown of the PEM 
approximation.
This is not surprising since we now reach the photonic crystal limit 
and the internal structure of the meta-material's constituents must become
visible and no ``effective description'' should be possible anymore.

\medskip
\paragraph*{Momentum dependent parameters.}
In order to take into account the periodic structure of real meta-materials,
we considered the PEM model as a very simple explicit example 
of a periodic medium.
Instead of qualifying the specific geometry of the model used, 
we may alternatively introduce $k$-dependent effective parameters
$\mu(k,\omega)$, $\varepsilon(k,\omega)$ to describe the spatial 
distribution of the electromagnetic response in the PEM.
These $k$-dependent effective parameters completely 
characterize the effective medium model.
For the PEM model defined in Fig.~\ref{fig_model_1} we find 
by Fourier transformation 
$\varepsilon(k,\omega)=(2\pi)^{-1/2}\int dz\ \varepsilon(z,\omega)\,e^{ikz}$
the 
representation
\begin{eqnarray}
\lefteqn{\varepsilon(k,\omega) \ =\ } && \\
&&
\sqrt{2\pi}\ 
\varepsilon_\mathrm{core}(\omega)\
\frac{e^{ik(L-b)}-e^{ika}}{ikL}\ 
\sum_{m\in{\mathbb Z}}\,\delta(k-\frac{2\pi m}{L}),\quad
\nonumber
\end{eqnarray}
and correspondingly for $\mu(k,\omega)$.
As a generalization of the periodic medium approximation we could further ask,
which arbitrary $k$-dependence of $\varepsilon(k,\omega)$ and $\mu(k,\omega)$,
ie.~which spatial distribution of the effective material parameters,
describes a given meta-material best. 
Although this might be desirable as a descriptive tool for the engineering of
meta-materials, it is clearly beyond the scope of this paper.

\medskip
\paragraph*{What is the actual left-handed band?}
Finally we want to comment on the actual extent of the left-handed interval 
of the LHM as it concerns experiments and applications.
Obviously, the bands with $\mathrm{Re}\ n_\mathrm{eff}<0$ retrieved via HEM
and PEM approximation differ considerably in width (Fig.~\ref{fig_lhm_1})
which raises the question, where to expect the left-handed behavior.
We argue, that the correct region is given by the HEM approximation.
If there is a length-independent HEM approximation, the scattering behavior
of the meta-material can be described assuming plain-wave solutions inside 
the homogeneous unit cell.
This plainwaves will possess a wavevector $q$ related to the vacuum-wavevector
$k$ by the retrieved negative index of refraction, $q=n_\mathrm{hem}(\omega)\,k$.
They coincide with the non-periodic factor of the Bloch-waves describing the
periodic medium, ie.~coincide with the Bloch-waves at the edges of the unit cells
in the meta-material.
Therefore a (usually damped) plainwave with negative phase-velocity will exist 
inside the meta-material whenever $\mathrm{Re}\ n_\mathrm{hem}<0$.
This interpretation is also supported by experimental measurements\cite{Parazzoli03}.
Note that this frequency interval is wider than the interval with simultaneously
negative $\mathrm{Re}\ \varepsilon_\mathrm{hem}$ and 
$\mathrm{Re}\ \mu_\mathrm{hem}$.
The PEM approximation indicates that the isolated local response of the SRR and
wire, without the effects of periodicity, would lead to a much smaller left-handed 
band (compare the behavior of the low-frequency SRR, Figs.~\ref{fig_srr_low_1},
\ref{fig_srr_low_2}).
The modifications of the generic response of SRR and wire by the band-structure,
in particular by the emergence of periodicity band-gaps,
arising from the inherent periodicity of the meta-material, greatly enhances
the width of the negative index band in meta-materials which see strong artifacts
from the periodicity. 
This concerns virtually all published meta-materials with a vacuum-wavelength to
unit cell length ratio smaller than approximately ten.
At much lower frequency, truly effective homogeneous behavior will emerge,
the periodicity band-gaps disappear, and HEM and PEM description coincide.

\medskip
\paragraph*{Geometry of the PEM model.}
As we demonstrated in section \ref{section:pem}, 
for the quasi one-dimensional scattering problem of a system 
comprised of an integral number of unit cells in propagation direction, 
we can exactly describe any HEM by a family of PEM, 
parametrized by the geometry ${\mathcal G}=(n_a,n_d,n_b)$ 
of the periodic medium model (cf.~Fig.~\ref{fig_model_1}), 
in terms of effective parameters
$\mu_{\mathrm{pem}[{\mathcal G}]}(\omega)$ and
$\varepsilon_{\mathrm{pem}[{\mathcal G}]}(\omega)$.
For a simulated meta-material, in general all effective parameters, 
$\mu_\mathrm{hem}(\omega)$ and $\varepsilon_\mathrm{hem}(\omega)$ as well as
$\mu_{\mathrm{pem}[{\mathcal G}]}(\omega)$ and
$\varepsilon_{\mathrm{pem}[{\mathcal G}]}(\omega)$,
will show ``unphysical'' behavior caused by the internal spatial structure 
of the unit cell and the periodicity of the meta-materials.
If for a real meta-material the local electromagnetic behavior of the 
constituents can be abstracted from their geometrical form,
approximated by simple resonant and plasmonic response 
functions $\mu_\mathrm{SRR}(\omega)$ and $\varepsilon_\mathrm{wire}(\omega)$, 
respectively, 
and separated from the effects of the periodicity (ie.~band-structure),
then there is a particular geometry of the PEM which approximates 
physical parameters
$\mu_{\mathrm{pem}[{\mathcal G}]}(\omega)\approx \mu_\mathrm{SRR}(\omega)$
and 
$\varepsilon_{\mathrm{pem}[{\mathcal G}]}(\omega)
 \approx \varepsilon_\mathrm{wire}(\omega)$
without the usual artifacts discussed in this paper.
In our simulations the best such approximation was obtained if the lattice PEM
containing a single plain of scatterers in the unit cell, 
ie.~for a geometry ${\mathcal G}=(5,1,4)$.

\section{\label{section:conclusion} Conclusion}

We have investigated the influence of the inherent periodic structure always
present in meta-materials which are build from the repetition of a single
unit cell on the effective medium approximation.
It has been shown analytically that all the previously observed violations 
of the anticipated effective medium behavior of the (single-ring) SRR and LHM 
involving a single magnetic and a single electric resonance can be explained
in term of the periodic structure: A very simple stratified periodic medium
model involving slabs of vacuum alternating with slabs of a homogeneous
material with simple resonant $\mu(\omega)$ and $\varepsilon(\omega)$
can reproduce all the artifacts like resonance/anti-resonance coupling in
$\mu_\mathrm{hem}(\omega)$ and $\varepsilon_\mathrm{hem}(\omega)$, negative
imaginary parts in either $\varepsilon$ or $\mu$, truncated, misshapen 
resonances, additional band gaps and the complicated high-frequency behavior
found in the HEM approximation of numerically simulated SRR arrays and LHM, 
but also meta-materials build of continuous wires and of cut-wires.
In good approximation, the effective behavior can be decomposed into a 
effective behavior of the constituents of the meta-material and an explicit
contribution of the periodicity.
Remarkably, the average contribution of the constituents like the single
split-ring behaves much like expected from the assumed homogeneous medium 
picture, which can only be justified in the low-frequency limit where the
wavelength inside the structure is large compared to its geometrical size,
up to frequencies where the vacuum wave length becomes comparable to the
size of the unit cell.
This allows a more reliable effective description and interpretation 
of real meta-materials in terms of a periodic effective medium (PEM) instead
of the conventional homogeneous effective medium (HEM) with all the
hard to understand artifacts.
The effects caused by the periodicity are generic, they do qualitatively
not depend on the particular geometry chosen for the meta-material and
universally apply to SRR, LHM, continuous wire and cut-wire materials. 
Obviously, the impact of the periodicity is noticeable 
throughout the range of the $\lambda_0/L$ ratio $\approx 5\ldots10$ 
(ie.~vacuum wave length / unit cell length)
actually found in published simulations and 
experiments for left-handed and related meta-materials.
Our simulations indicate that an unencumbered homogeneous effective medium
behavior, though reachable in the low-frequency limit, would require a
$\lambda_0/L$ ratio in the order of 30 or larger which is geometrically 
not easy to obtain in real samples.
We investigated the difference between the continuum and the lattice 
formulation of the PEM approximation and found the latter to be better 
suited for application to our numerically simulated scattering data 
for real meta-materials obtained with a lattice-TMM implementation.

The PEM approximation may provide an valuable tool to understand the
various features observed in the scattering spectra of real meta-materials 
in experiments and simulations. 
In the present paper we essentially discussed the vicinity of the magnetic
resonance of the SRR.
Further work shall emphasize on the frequencies above the magnetic resonance
$\omega_m$ including the cut-wire resonance $\omega_e$.

We expect the impact of the meta-material's periodicity to be noticeable 
also in higher dimensional structures. Because the unit cells of those 
structures tend to be more complicated including couplings of SRRs in the 
different directions the separation of the ``real'' effective response 
of the constituents and the structures produced by the periodicity 
constitutes an even more immanent issue for understanding.

\section{Acknowledgements}

This work was partially supported by Ames Laboratory 
(Contract number W-7405-Eng-82). 
Financial support of EU$\underline{~~}$FET project DALHM, 
NSF (US-Greece Collaboration), 
and DARPA (Contract number MDA972-01-2-0016) are also acknowledged.
PM thanks APVT (Grant number 51-021602) for partial financial support.

\bibliographystyle{apsrev}

\begin{thebibliography}{99}
\bibitem{ARCCOS}
We define the principal branch of $\arg(z)$ as $-\pi < \arg\,z \le \pi$ 
with a branch-cut along the negative real axis, 
continuous from the second quadrant.
Then the principal branches of 
$\sqrt{z} = |z|^{1/2}\exp(i \arg(z)/2 )$ and $\log(z) = \log |z| + i\arg(z)$
can be defined with the same branch-cut and continuity from above.
Using $2\cos(z)=\exp(iz)+\exp(-iz)$
we obtain the principal branch 
$\arccos(z) = -i \log(z+i\sqrt{1-z^2})$ with a branch-cut on the real axis
for $z\in(1,\infty)$, continuous from below,
and $z\in(-\infty,-1)$, continuous from above.
\end{thebibliography}

\end{document}